# GRAVITATIONAL WAVE DETECTION IN SPACE*

Wei-Tou Ni

*School of Optical Electrical and Computer Engineering,*
*University of Shanghai for Science and Technology,*
*516, Jun Gong Rd., Shanghai 200093, China*

*and*

*Kavli Institute for Theoretical Physics China, CAS, Beijing 100190, China*
weitouni@163.com; weitou@gmail.com

Gravitational wave (GW) detection in space is aimed at low frequency band (100 nHz – 100 mHz) and middle frequency band (100 mHz – 10 Hz). The science goals are the detection of GWs from (i) Supermassive Black Holes; (ii) Extreme-Mass-Ratio Black Hole Inspirals; (iii) Intermediate-Mass Black Holes; (iv) Galactic Compact Binaries and (v) Relic GW Background. In this paper, we present an overview on the sensitivity, orbit design, basic orbit configuration, angular resolution, orbit optimization, deployment, time-delay interferometry and payload concept of the current proposed GW detectors in space under study. The detector proposals under study have arm length ranging from 1000 km to $1.3 \times 10^9$ km (8.6 AU) including (a) Solar orbiting detectors -- ASTROD-GW (ASTROD [Astrodynamical Space Test of Relativity using Optical Devices] optimized for GW detection), BBO (Big Bang Observer), DECIGO (DECi-hertz Interferometer GW Observatory), e-LISA (evolved LISA [Laser Interferometer Space Antenna]), LISA, other LISA-type detectors such as ALIA, TAIJI etc. (in Earth-like solar orbits), and Super-ASTROD (in Jupiter-like solar orbits); and (b) Earth orbiting detectors -- ASTROD-EM/LAGRANGE, GADFLI/GEOGRAWI/g-LISA, OMEGA and TIANQIN.

## 1. Introduction

Gravitational Wave (GW) detection has been a focused research subject for some time. With the announcement of LIGO direct GW detection [1, 2], we are fully ushered into the age of GW astronomy. Second-generation ground-based interferometers are being upgraded/completed for GW detection in the high-frequency band (10–100 kHz; see Ref.s [3-5] for a complete spectral classification of GWs) [6]. Observational data from Pulsar Timing Arrays (PTAs) are being accumulated for the first GW detection in the very low frequency band (300 pHz–100 nHz) [7]. Collaborations working on Cosmic Microwave Background (CMB) observations are actively pushing their sensitivities

*This paper is to be published also as chapter 13 of the book "*One Hundred Years of General Relativity*: *From Genesis and Empirical Foundations to Gravitational Waves, Cosmology and Quantum Gravity*", edited by Wei-Tou Ni (World Scientific, Singapore, 2016).

further for detecting imprints of primordial GWs in the Hubble frequency band (1 aHz–10 fHz) on B-mode polarizations [8]. LISA (Laser Interferometric Space Antenna)[9] Pathfinder[10] launched on 3 December 2015 has successfully demonstrated the drag-free technology[11] for space detection of GWs in the middle and low frequency band (0.1 Hz–10 Hz; 100 nHz–0.1 Hz). The activities are mounting in this centennial year (2015-2016) of the establishment of general relativity.

With the invention of lasers in 1960, the implementation of satellite laser ranging and lunar laser ranging in 1960s and the development of drag-free navigation for geodesy in 1970s, concept of laser interferometry in space for GW detection were developed in 1980s. The first public proposal on space interferometers for GW detection was presented at the Second International Conference on Precision Measurement and Fundamental Constants (PMFC-II), 8–12 June 1981, in Gaithersburg [12,13]. In this seminal proposal, Faller and Bender raised possible GW mission concepts in space using laser interferometry. Two basic ingredients were addressed — drag-free navigation for the reduction of perturbing forces on the spacecraft (S/C) and laser interferometry for the sensitivity of measurement. LISA-like S/C orbit formation was reached in 1985 in the proposal Laser Antenna for Gravitational-radiation Observation in Space (LAGOS).[14] A schematic of LISA-type orbit configuration is shown in Fig. 1. It is natural for people like Bender and Faller working in lunar laser ranging and measuring free-fall acceleration using interferometry to propose such an experiment. In fact, test mass free fall inside a falling shroud in vacuum in the interferometric measurement of the Earth's gravitational acceleration can be considered as a passive drag-free navigation device.[15] The discrepancy in the absolute gravimeter comparison at the BIPM (Bureau International des Poids et Mesures) is partially resolved using correction to interferometric measurements of absolute gravity arising from the finite speed of light.[16] In the S/C tracking, the finite velocity of light has always been incorporated. Both the test mass for GW missions and the test mass of interferometric gravimeter can be regarded as freely falling objects in the solar system and tracked using astrodynamical equation. Thus, we see the interplay among space geodesy, Galileo Equivalence Principle (Universality of Free Fall) experiments in space and GW detection missions. Recent development for a GRACE follow-on mission SAGM (Space Advanced Gravity Measurements),[17] TEPO[18] (testing the equivalence principle with optical readout in space) and TIANQIN[19] (a space-borne GW detector) can be considered as such an example.

A big step for the GW detection in space is the 1993 ESA M3 Assessment study of LISA and later recommendation as the third cornerstone of "Horizon 2000 Plus". After 2000, LISA became a joint ESA–NASA mission until the 2011 NASA withdrawal. In 1998, LISA Pathfinder was selected as the second of the European Space Agency's Small Missions for Advanced Research in Technology (SMART) to develop and to test the demanding drag-free technology. At this occasion of Centennial Celebration of General Relativity, ESA has successfully launched the LISA Pathfinder on a Vega rocket from Europe's spaceport in Kourou, French Guiana on 3 December, 2015, and has successfully demonstrated the drag-free technology[11] for observing GWs from space. Based on the ongoing technological development for LISA Pathfinder, ESA has sponsored a technology reference study (completed in 2008) for the fundamental physics explorer as a common bus for fundamental physics missions.[20] NGO (New Gravitational-wave Observatory)/eLISA (evolved LISA),[21] down-scaled from 5 million km to 1 million km arm length, was proposed in 2011 to accommodate the budget change and received excellent evaluation. In November 2013, ESA announced the selection of the Science Themes for the L2 and L3 launch opportunities – the "Hot and Energetic Universe" for L2 and "The Gravitational Universe" for L3.[22] ESA L3 mission is likely to have a launch opportunity in 2034.[22] Since eLISA/NGO GW mission concept is the major candidate at this time and it takes one year to transfer to the science orbit, a starting time for science phase is likely in 2035. Since 2035 is still 20 years away, it is not yet the time to freeze the specific mission concept. At present a comparison of laser measurement technology and atom interferometry is underway in ESA.

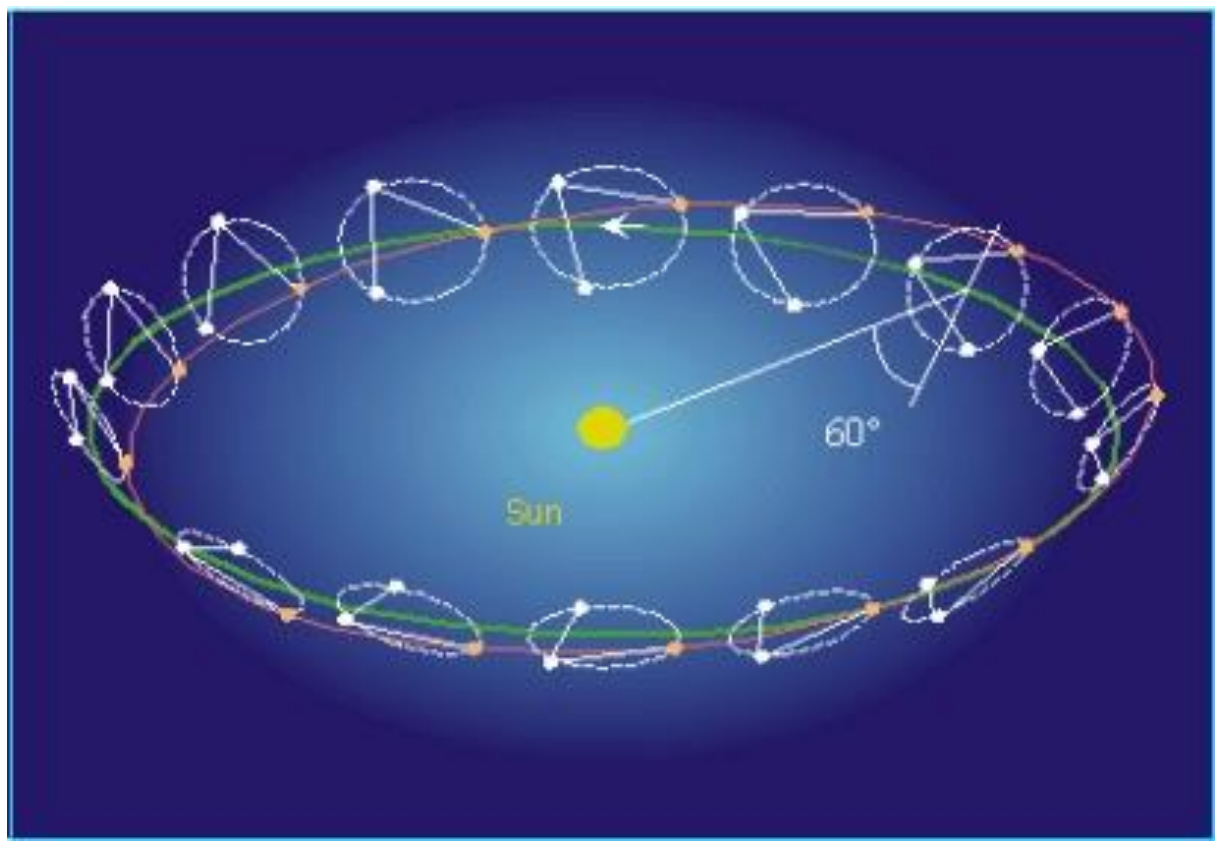

**Fig.1.** Schematic of LISA-type orbit configuration in Earth-like solar orbit.[9]

The general concept of Astrodynamical Space Test of Relativity using Optical Devices (ASTROD) is to have a constellation of drag-free S/Cs navigate through the solar system and range with one another using optical devices to map the solar-system gravitational field, to measure related solar-system parameters, to test relativistic gravity, to observe solar g-mode oscillations and to detect GWs. A baseline implementation of ASTROD was proposed in 1993 and has been under concept and laboratory studies since then.[23-30] In 1996, ASTROD I (Mini-ASTROD) with one S/C ranging with ground stations was proposed for testing relativistic gravity and mapping the solar system.[23] The mission study shows that the precision of testing relativistic gravity in the solar system is achievable to $10^{-9}$–$10^{-8}$ in terms of Eddington parameter $\gamma$, which is more than three orders of improvement over the present precision, with accompanying improvement in other aspects of relativistic gravity.[31-35] Early in 2009, responding to the call for GW mission studies of CAS (Chinese Academy of Sciences), a dedicated mission concept ASTROD-GW (ASTROD optimized for Gravitational Wave [GW] detection) for GW detection with 3 S/C (spacecraft) orbiting near Sun-Earth Lagrange points L3, L4 and L5 respectively with nominal arm length of 260 million km was proposed and studied.[3,36-40] A schematic of ASTROD-GW orbit configuration with inclination is shown in Fig. 2.[3,41]

Before the ASTROD-GW proposal, Super-ASTROD which was proposed in 1996[23] with S/C's in Jupiter-like orbits was studied as a dual mission for GW measurement and for cosmological model/relativistic gravity test in 2008.[42] With the proposal of ASTROD-GW, the baseline GW configuration of Super-ASTROD makes 3 out of 4-5 S/C orbiting near Sun-Jupiter Lagrange points L3, L4 and L5 respectively. For the possibility of a down scaled version of ASTROD-GW mission, the ASTROD-EM with the orbits of 3 S/C near Earth-Moon Lagrange points L3, L4 and L5 respectively has been under study.[43]

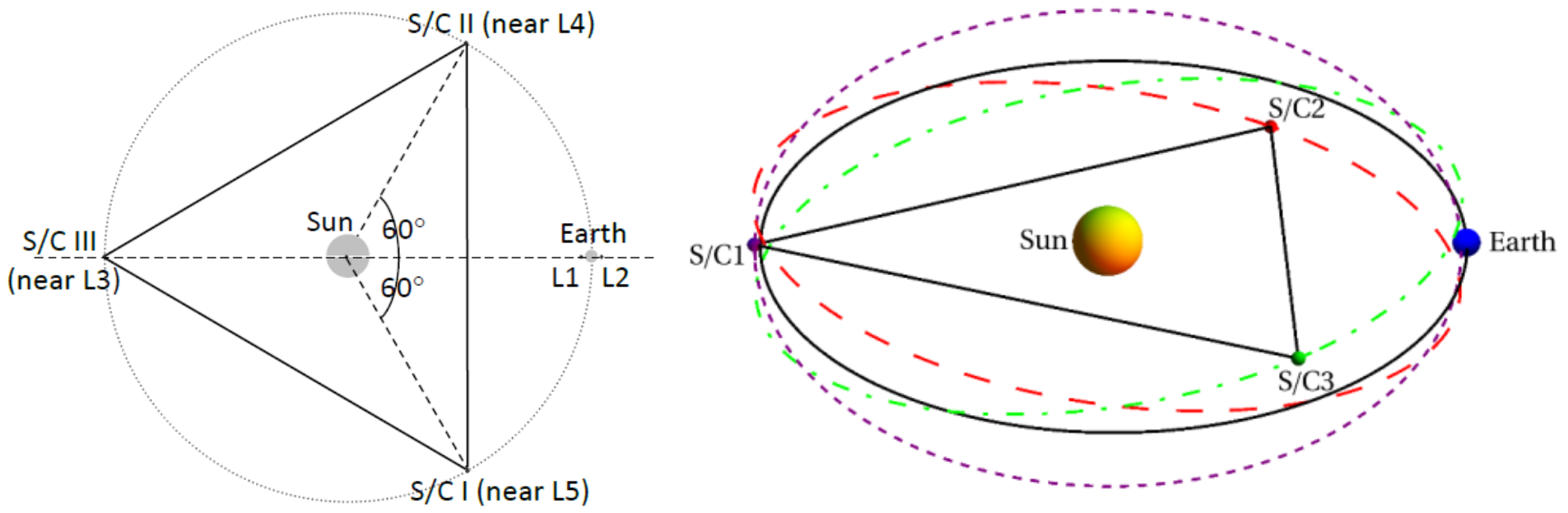

**Fig.2.** Schematic of ASTROD-GW orbit configuration with inclination. Left, projection on the ecliptic plane; Right, 3-d view with the scale of vertical axis multiplied tenfold.[3,41]

DECi-hertz Interferometer GW Observatory (DECIGO)[44] was proposed in 2001 with the aim of detecting GWs from early universe in the middle frequency observation band between the terrestrial band and the low frequency band of other space GW detectors. It will use a Fabry-Perot method (instead of a delay line method) as in the ground interferometers but with a 1000 km arm length. As a LISA follow-on, BBO (Big Bang Observer)[45] with arm length 50,000 km was proposed in the United States with a similar goal. A likely version of DECIGO/BBO is to have 12 S/Cs with correlated detection. They will be used for the direct measurement of the stochastic GW background by correlation analysis.[46] 6S/C-ASTROD-GW with two sets of ASTROD-GW has also been considered to possibly explore the relic GWs in the lower part of the low frequency band.[39,40] ALIA[47] of arm length 500,000 km was proposed as a less-ambitious LISA follow-on. TAIJI (also called ALIA descope)[48] of arm length 3 million km has also been proposed and under study with the main goal of detecting intermediate mass black hole binaries at high redshift.

After the end in 2011 of ESA-NASA partnership for flying LISA, NASA solicited "Concepts for the NASA Gravitational Wave Mission" proposals on 27 September 2011 for study of low cost GW missions (http://nspires.nasaprs.com/external/). gLISA/GEOGRAWI[49-51] (geosynchronous LISA / GEOstationary GRAvitational Wave Interferometer), GADFLI[52] (Geostationary Antenna for Disturbance-Free Laser Interferometry), and LAGRANGE[53] (Laser Gravitational-wave Antenna at Geo-lunar Lagrange points) was proposed and OMEGA[54,55] (Orbiting Medium Explorer for Gravitational Astronomy) re-emerged. OMEGA of arm length 1 million km was first proposed as a low-cost alternative to LISA in the 1990s. An artist's conception of the OMEGA mission configuration is shown in Fig. 3. In China, a GW mission in Earth orbit called TIANQIN[56] of arm length 110,000 km has been proposed and under study.

Table 1 lists the orbit configuration, arm length, orbit period, S/C number, acceleration noise and laser metrology noise of various GW space mission proposals. Fig.'s 4-6 show respectively the strain psd (power spectral density) amplitude $[S_h(f)]^{1/2}$ versus frequency plot, the characteristic strain $h_c$ versus frequency plot and the normalized GW spectral energy density $\Omega_{gw}$ versus frequency plot for various GW detectors and sources in the low-frequency band and middle frequency band. The characteristic strain $h_c$, the strain psd amplitude $[S_h(f)]^{1/2}$ and the normalized GW spectral energy density $\Omega_{gw}$ are related as follows:

$$h_c(f) = f^{1/2} [S_h(f)]^{1/2}; \ \Omega_{gw}(f) = (2\pi^2/3H_0^2)\, f^3\, S_h(f) = (2\pi^2/3H_0^2)\, f^2\, h_c^2(f). \quad (1)$$

Detailed accounts and explanations of Fig.'s 4-6 are given in Sec.'s 3-6 and in Ref. [5]. A large part of these figures are taken from the corresponding low frequency band and middle frequency band of Fig.'s 2-4 in Ref. [5].

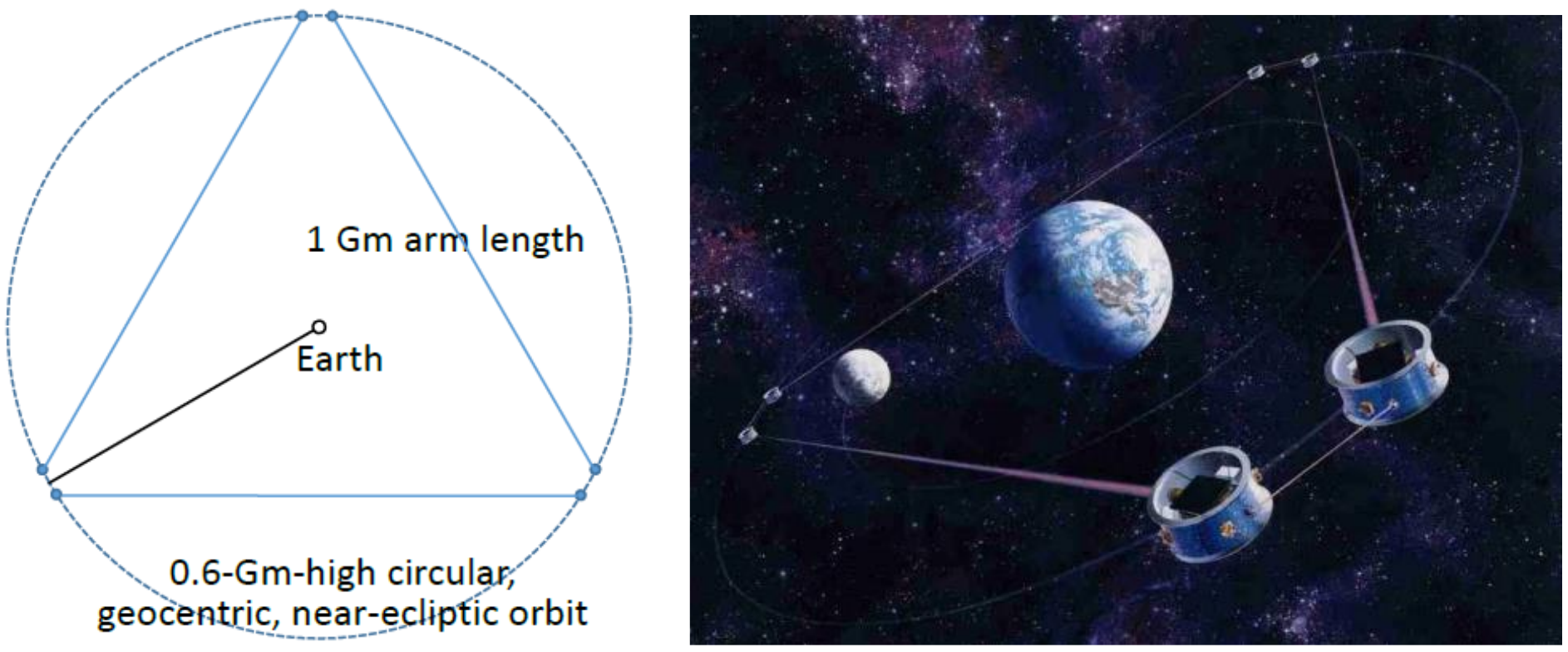

**Fig. 3.** Schematic (left) and artist's conception (right) of the OMEGA mission configuration.[55]

Table 1. A Compilation of GW Mission Proposals

| Mission Concept | S/C Configuration | Arm length | Orbit Period | S/C # | Acceleration noise [fm/s²/Hz^1/2] | laser metro-logy noise [pm/Hz^1/2] |
|---|---|---|---|---|---|---|
| *Solar-Orbit GW Mission Proposals* | | | | | | |
| LISA[9] | Earth-like solar orbits with 20° lag | 5 Gm | 1 year | 3 | 3 | 20 |
| eLISA[21] | Earth-like solar orbits with 10° lag | 1 Gm | 1 year | 3 | 3 | 12 (10) |
| ASTROD-GW[36-40] | Near Sun-Earth L3, L4, L5 points | 260 Gm | 1 year | 3 | 3 | 1000 |
| Big Bang Observer[45] | Earth-like solar orbits | 0.05 Gm | 1 year | 12 | 0.03 | $1.4 \times 10^{-5}$ |
| DECIGO[44] | Earth-like solar orbits | 0.001 Gm | 1 year | 12 | 0.0004 | $2 \times 10^{-6}$ |
| ALIA[47] | Earth-like solar orbits | 0.5 Gm | 1 year | 3 | 0.3 | 0.6 |
| TAIJI (ALIA-descope)[48] | Earth-like solar orbits | 3 Gm | 1 year | 3 | 3 | 5-8 |
| Super-ASTROD[42] | Near Sun-Jupiter L3, L4, L5 points (3 S/C), Jupiter-like solar orbit(s)(1-2 S/C) | 1300 Gm | 11 year | 4 or 5 | 3 | 5000 |
| *Earth-Orbit GW Mission Proposals* | | | | | | |
| OMEGA[54,55] | 0.6 Gm height orbit | 1 Gm | 53.2 days | 6 | 3 | 5 |
| gLISA/GEOGRAWI[49-51] | Geostationary orbit | 0.073 Gm | 24 hours | 3 | 3, 30 | 0.3, 10 |
| GADFLI[52] | Geostationary orbit | 0.073 Gm | 24 hours | 3 | 0.3, 3, 30 | 1 |
| TIANQIN[56] | 0.057 Gm height orbit | 0.11 Gm | 44 hours | 3 | 1 | 1 |
| ASTROD-EM[43] | Near Earth-Moon L3, L4, L5 points | 0.66 Gm | 27.3 days | 3 | 1 | 1 |
| LAGRANGE[53] | Earth-Moon L3, L4, L5 points | 0.66 Gm | 27.3 days | 3 | 3 | 5 |

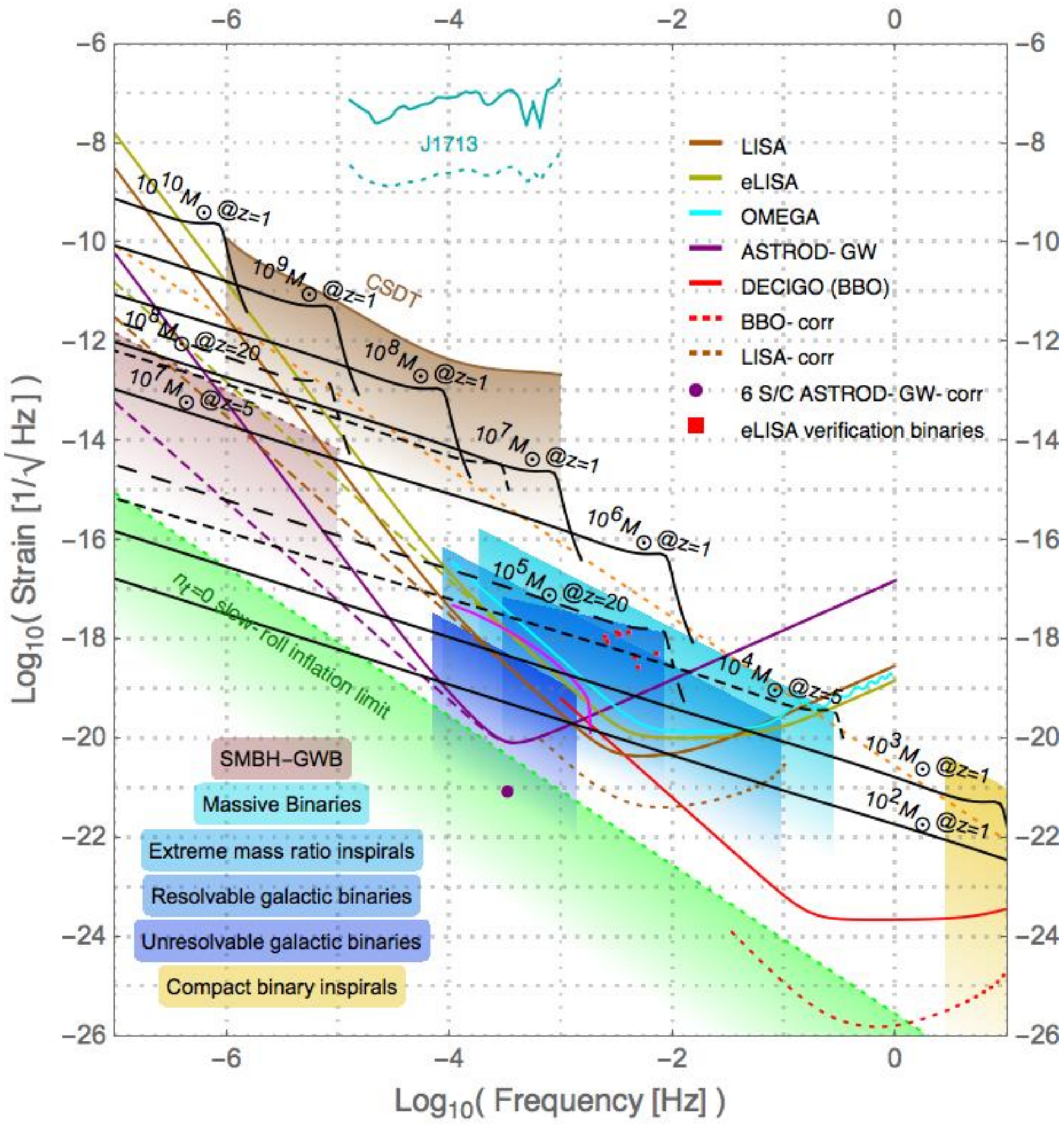

Fig. 4. Strain power spectral density (psd) amplitude vs. frequency for various GW detectors and GW sources. The black lines show the inspiral, coalescence and oscillation phases of GW emission from various equal-mass black-hole binary mergers in circular orbits at various redshift: solid line, $z$ = 1; dashed line, $z$ = 5; long-dashed line $z$ = 20. See text for more explanation. [CSDT: Cassini Spacecraft Doppler Tracking; SMBH-GWB: Supermassive Black Hole-GW Background.]

In the following section, we discuss the link of gravity (including GW) with orbit observations/experiments in the solar system. In section 3, we review the methods and the most recent experimental results of radio Doppler spacecraft tracking. In section 4, we explain the basic principle of laser-interferometric space mission for GW detection.

In section 5, we address the sensitivity spectra and review basic noises. In section 6, we discuss the scientific goals of GW space missions. In Sec. 7, we address the basic orbit design using eLISA and ASTROD-GW as concrete examples. In Sec. 8, we discuss the orbit design and orbit optimization using ephemerides. In Section 9, we discuss the deployment of spacecraft to various positions of Earth-like solar orbit, their propellant ratios and the total mass requirements. In Sec. 10, we discuss time delay interferometry. In Sec. 11, we discuss the payload. In Sec. 12, we summarize the paper and present an outlook.

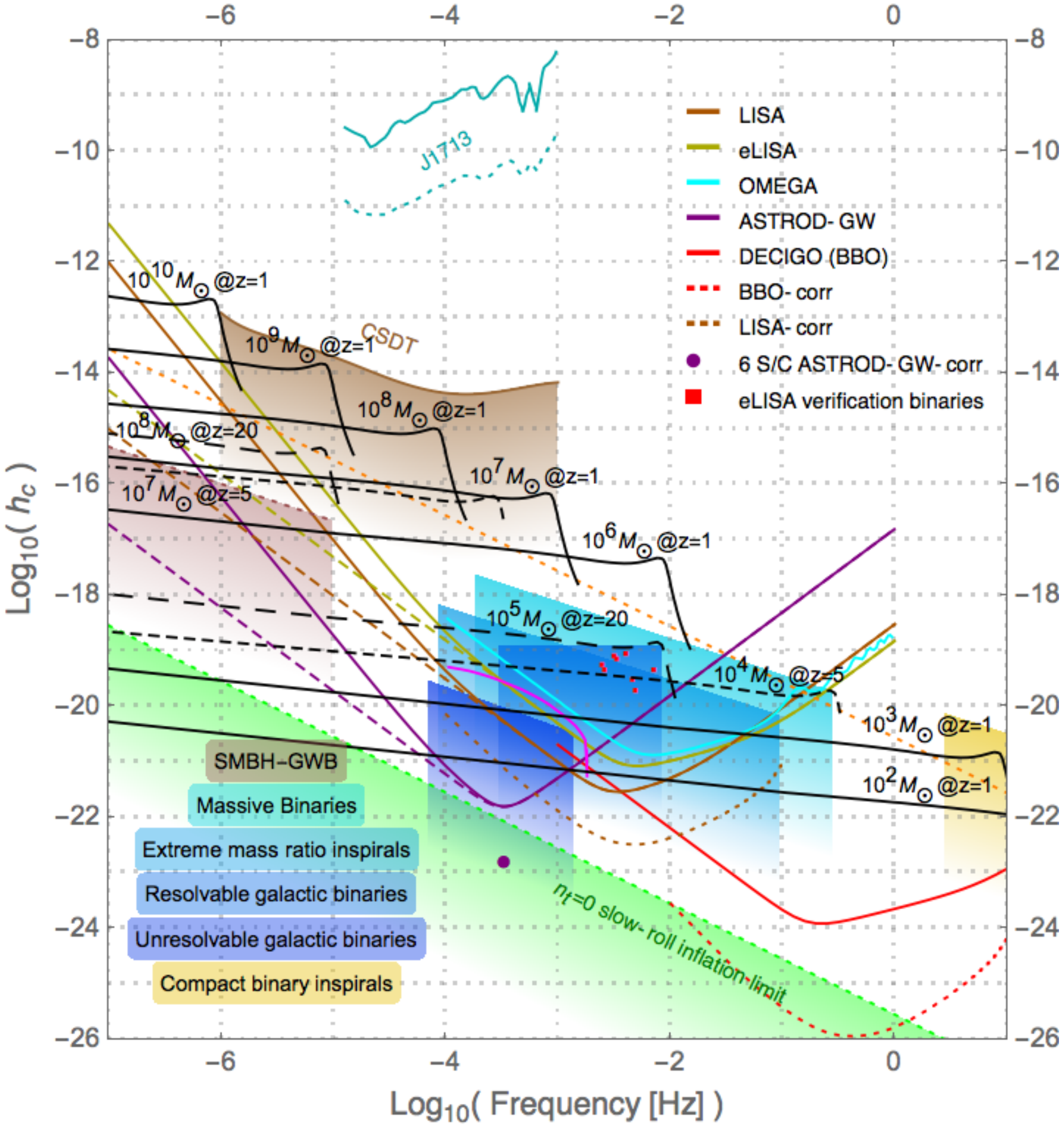

Fig. 5. Characteristic strain $h_c$ vs. frequency for various GW detectors and sources. The black lines show the inspiral, coalescence and oscillation phases of GW emission from various equal-mass black-hole binary mergers in circular orbits at various redshift: solid line, $z$ = 1; dashed line, $z$ = 5; long-dashed line $z$ = 20. See text for more explanation. [CSDT: Cassini Spacecraft Doppler Tracking; SMBH-GWB: Supermassive Black Hole-GW Background.]

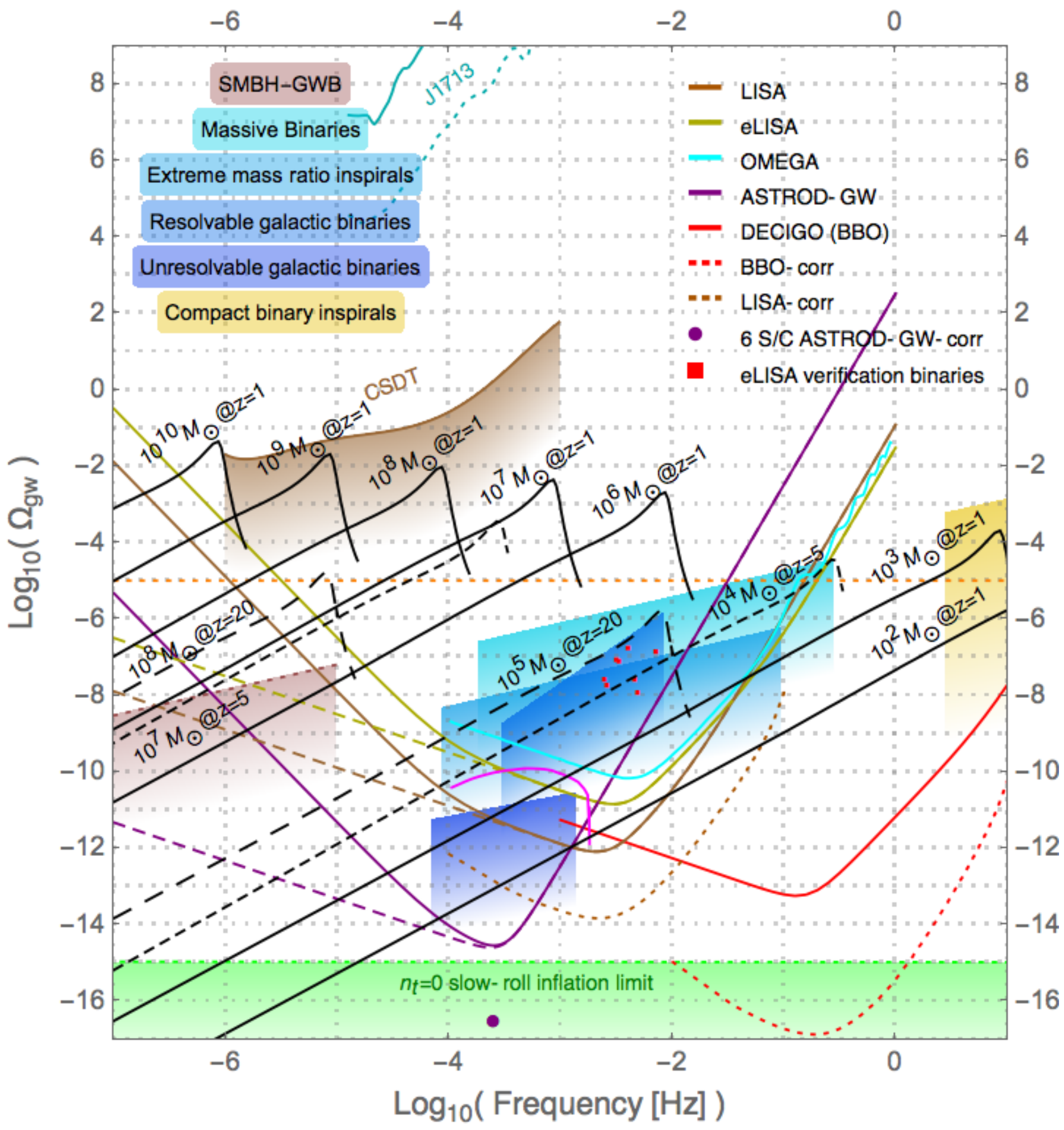

Fig. 6. Strain power spectral density (psd) amplitude vs. frequency for various GW detectors and GW sources. The black lines show the inspiral, coalescence and oscillation phases of GW emission from various equal-mass black-hole binary mergers in circular orbits at various redshift: solid line, $z$ = 1; dashed line, $z$ = 5; long-dashed line $z$ = 20. See text for more explanation. [CSDT: Cassini Spacecraft Doppler Tracking; SMBH-GWB: Supermassive Black Hole-GW Background.]

## 2. Gravity and Orbit Observations/Experiments in the Solar-System

Historically the orbit and gravity observations/experiments in the solar-system have been important resources for the development of fundamental physical laws as the precision and accuracy are improved. It is so for both the developments of Newtonian world system and Einstein's general relativity.[57-59] With the eminent improvement for orbit and gravity measurements pending, we are in a historical epoch for a great stride in the testing and development of fundamental laws. The gravitational field in the solar system is determined by three factors: the dynamic distribution of matter in the solar system; the dynamic distribution of matter outside the solar system (galactic, cosmological, etc.) and GWs propagating through the solar system. Different relativistic/cosmological theories of gravity make different predictions of the solar-system gravitational field. Hence, precise measurements of the solar-system gravitational field test these relativistic theories, in addition to enabling GW observations, determination of the matter distribution in the solar-system and determination of the observable (testable) influence of our galaxy and cosmos. To measure the solar-system gravitational field, we measure/monitor distance between different natural and/or artificial celestial bodies. In the solar system, the equation of motion of a celestial body or a spacecraft is given by the astrodynamical equation

$$\mathbf{a} = \mathbf{a}_{\mathrm{N}} + \mathbf{a}_{\mathrm{1PN}} + \mathbf{a}_{\mathrm{2PN}} + \mathbf{a}_{\mathrm{Gal\text{-}Cosm}} + \mathbf{a}_{\mathrm{GW}} + \mathbf{a}_{\mathrm{non\text{-}grav}}\,, \qquad (2)$$

where $\mathbf{a}$ is the acceleration of the celestial body or spacecraft, $\mathbf{a}_{\mathrm{N}}$ is the acceleration due to Newtonian gravity, $\mathbf{a}_{\mathrm{1PN}}$ the acceleration due to first post-Newtonian effects, $\mathbf{a}_{\mathrm{2PN}}$ the acceleration due to second post-Newtonian effects, $\mathbf{a}_{\mathrm{Gal\text{-}Cosm}}$ the acceleration due to Galactic and cosmological gravity, $\mathbf{a}_{\mathrm{GW}}$ the acceleration due to GWs, and $\mathbf{a}_{\mathrm{nongrav}}$ the acceleration from all non-gravitational origins.[3] Distances between spacecraft depend critically on the solar-system gravity (including gravity induced by solar oscillations), underlying gravitational theory and incoming GWs. A precise measurement of these distances as a function of time will enable the cause of variation to be determined.

Ideally it would be desirable to have a constellation of drag-free spacecraft navigate through the solar system and range with one another using optical devices (or other sensitive devices) to map the solar-system gravitational field, to measure related solar-system parameters, to test relativistic gravity, to observe solar g-mode oscillations, and to detect GWs.[3,60] Practically, certain orbit configurations are good for testing relativistic gravity; certain configurations are good for measuring solar parameters; certain are good for detecting gravitational waves. These factors are integral part of mission designs for various purposes.[3,60]

To test relativistic gravity, the spacecraft needs to go into inner solar orbit where the solar gravity is stronger or to send signals passing near the solar limbs to get stronger influence from solar gravity. ASTROD I during the superior solar conjunctions to measure the Shapiro delay of light and with continuous laser ranging of 1 mm accuracy to improve the determination of relativistic parameters is such a mission proposal.[31-33] BepiColombo to be launched in 2017 is an ESA-JAXA mission under implementation.[61,62] One of its goals of radio science is to test relativistic gravity. In determining its orbit about Mercury, it will indirectly find the motion of the center of mass of Mercury with an accuracy several orders of magnitude better than what is possible by radar ranging to its surface. This is a good opportunity to measure Mercury's perihelion advance and the Shapiro time delay, and to improve on the other post-Newtonian parameters by a couple of orders of magnitude.[63]

To measure or to improve solar and planetary parameters, the spacecraft needs to go near the measured body or to have supreme sensitivity. NEAR (Near Earth Asteroid Rendezvous Mission: determined the mass $(6.687 \pm 0.003) \times 10^{18}$ gm and density $2.67 \pm 0.03$ gm/cm$^3$ of asteroid 433 Eros, its lower order gravitational-harmonics, and its rotation state using ground-based Doppler and range tracking of the NEAR spacecraft orbiting Eros together with images of the asteroid's surface landmarks),[64] MESSENGER (MErcury Surface, Space ENvironment, GEochemistry, and Ranging: entered orbit around Mercury on March 18, 2011, deorbited as planned, and impacted the surface of Mercury on April 30, 2015.[65] During this period, MESSENGER measured the gravity of Mercury and the state of the planetary core by utilizing the spacecraft's positioning data.) and ASTROD I (a mission proposal having a Venus swing-by for gravity assistance and for improved measurement of Venus gravity/multipole moments, with laser ranging of accuracy about 1 mm for improvement on the parameter determination of planets and asteroids)[31-33] are such examples.

For laser-interferometric GW detection without fast Doppler tracking (e.g., using optical combs), nearly equal arm lengths are required; LISA-like mission concepts and ASTROD-GW-like mission concepts are examples.

## 3. Doppler Tracking of Spacecraft

Radio Doppler tracking of spacecraft in a space mission can be used to constrain (or detect) the level of low-frequency GWs. The separated test masses of this GW detector are the Doppler tracking radio antenna on Earth and a distant S/C. Doppler tracking measures relative distance change. From these measurements, GWs can be detected or constrained. In 1967, Braginsky and Gertsenshtein[66] first proposed to use Doppler data of spacecraft tracking for GW searches. In 1971, Anderson[67] pursued this method of search with preexisting data. Davis[68] worked out the GW response of Doppler tracking for special cases in 1974; Estabrook and Walquist[69] analyzed the effect of GWs passing through the line of sight of S/C on the Doppler tracking frequency measurements in general in 1975 (see also [70]).

In Doppler tracking of S/C, a highly stable master clock on Earth is used as a reference to control a monochromatic radio wave for transmitting to S/C (uplink). When S/C transponder receives the monochromatic radio wave, it phase-locks the local oscillator with or without a frequency offset and transponds the local oscillator signal back (to Earth station; downlink) coherently.

The one-way Doppler response $y(t)$ is defined as

$$y(t) \equiv \delta\nu/\nu_0 \equiv (\nu_1(t) - \nu_0)/\nu_0, \quad (3)$$

where $\nu_0$ is the frequency of emitted signal and $\nu_1$ is the frequency of received signal. Far from the GW sources as it is in the present experimental/observational situations, the plane wave approximation is valid. For weak plane waves propagating in the $z$-direction in general relativity, we have the following spacetime metric:

$$ds^2 = dt^2 - (\delta_{ij} + h_{ij}(ct - z))dx^i dx^j, \qquad |h_{ij}| << 1, \quad (4)$$

where Latin indices run from 1 to 3 and sum over repeated indices is assumed. Estabrook and Walquist[69,70] derived the one-way and two-way Doppler responses to plane GWs in weak field approximation (4) in the transverse traceless gauge in general relativity. Written in the notation of Armstrong, Estabrook and Tinto,[71] the formula for one-way Doppler response on board S/C 2 received from S/C 1 is

$$y(t) = (1 - \underline{\boldsymbol{k}} \cdot \underline{\boldsymbol{n}}) \, [\Psi(t - (1 + \underline{\boldsymbol{k}} \cdot \underline{\boldsymbol{n}})L) - \Psi(t)], \tag{5}$$

where $\underline{\boldsymbol{k}}$ [= $(\underline{k}^i) = (\underline{k}^1, \underline{k}^2, \underline{k}^3)$] is the unit vector in the GW propagation direction, $\underline{\boldsymbol{n}}$ [= $(\underline{n}^i) = (\underline{n}^1, \underline{n}^2, \underline{n}^3)$] the unit vector along the link from spacecraft 1 to spacecraft 2 and $L$ is the path length of the Doppler link. The function $\Psi(t)$ is defined as

$$\Psi(t) \equiv \underline{n}^i h_{ij}(t) \underline{n}^j \, / \, \{2[1 - (\underline{\boldsymbol{k}} \cdot \underline{\boldsymbol{n}})^2]\}. \tag{6}$$

With one-way Doppler response known, two-way and multiple way response can easily be written down. As noticed and derived by Tinto and da Silva Alves,[72] for GW solutions in any metric theories of gravity of the form (4), the Doppler response formula (5) and (6) are valid also.

Doppler tracking of the Viking S/C (S-band, 2.3 GHz),[73] the Voyager I S/C (S-band uplink + coherently transponded S-band and X-band (8.4 GHz) downlink),[74] Pioneer 10 (S band),[75] and Pioneer 11 (S band)[76] have been used for GW measurement and have given constraints on GW background in the low-frequency band.

The most recent measurements came from the Cassini spacecraft Doppler tracking (CSDT). Armstrong, Iess, Tortora, and Bertotti[77] used the Cassini multilink radio system during 2001–2002 solar opposition to derive improved observational limits on an isotropic background of low-frequency gravitational waves. The Cassini multilink radio system consists of a sophisticated multilink radio system that simultaneously receives two uplink signals at frequencies of X and Ka bands and transmits three downlink signals with X-band coherent with the X-band uplink, Ka-band coherent with the X-band uplink, and Ka-band coherent with the Ka-band uplink. X band is a standard deep space communication frequency band about 8.4 GHz; Ka band is another deep space communication frequency band about 32 GHz. Armstrong *et al.*[77] used the Cassini multilink radio system with higher frequencies and an advanced tropospheric calibration system to remove the effects of leading noises — plasma and tropospheric scintillation to a level below the other noises. The resulting data were used to construct upper limits on the strength of an isotropic background in the 1 μHz to 1 mHz band.[77] The characteristic strain upper limit curve labelled CSDT in Fig. 4 is a smoothed version of the curve in the Fig. 4 of Ref. [77]. The corresponding CSDT curves on the strain psd amplitude in Fig. 5 and the normalized spectral energy density in Fig. 6 are calculated using Eq. (1) for conversion. The minimal points on these curves are

$[S_h(f)]^{1/2} < 8 \times 10^{-13}$, at several frequencies in the 0.2-0.7 mHz band;
$h_c(f) < 2 \times 10^{-15}$, at frequency about 0.3 mHz;
$\Omega_{gw}(f) < 0.03$, at frequency 1.2 μHz. (7)

The GW sensitivity of spacecraft Doppler tracking could still be improved by 1-2 order of magnitude with a space borne optical clock on board.[78]

In the radio tracking of spacecraft the received frequency of the signals is tracked. Its integral is the phase. In the radio ranging of spacecraft the received phase of the signals is measured. The derivative of the phase is the frequency. For coherent transponding, the phase measured is basically a ranging up to an additive constant to be determined.

*Pulse laser ranging.* Another way to measure the range is by using pulse timing. This is what being done in satellite laser ranging and lunar laser ranging. For ranging through the Earth's atmosphere, the best way to find the atmospheric delay is to use two colors (two wavelengths) to measure the atmospheric delay and subtract it. The distance determination of satellite laser ranging with two colors (two wavelengths) has reached millimeter accuracy. With the newer generation of lunar laser ranging,[79,80] the accuracy of lunar distance determination has also reached millimeter accuracy. On board timing accuracy of 3 ps (0.9 mm) has already achieved by the T2L2 (Time Transfer by Laser Link) event timer onboard Jason 2 satellite.[81,82] Based on these developments, the one-way ranging technical capability over the whole solar system could have a millimeter accuracy. With this accuracy and extended ranges of 20 AU, the capability of probing the fundamental laws of spacetime and mapping the solar system gravity will be greatly enhanced.[32-35] For 1 mm out of 20 AU, the fractional uncertainty is $3 \times 10^{-16}$. It requires laser stability and clock accuracy to reach this level of fractional uncertainty; the accuracy is already achieved in the laboratory and will be available in space. ASTROD I[31-35] using a space borne precision clock has included as one of its goals GW sensitivity improvement of the Cassini spacecraft Doppler tracking by one order of magnitude. In fact, the fractional accuracies of optical clocks have already reached the $10^{-18}$ level. *When space optical clocks reach this level, pulse laser ranging together with drag-free technology will be an important alternative for detection of GWs in the lower part of low frequency band.*

The basic principle of spacecraft Doppler tracking, of spacecraft laser ranging, of space laser interferometers, and of pulsar timing arrays (PTAs) for GW detection are similar. In the development of GW detection methods, spacecraft Doppler tracking method and pulse laser ranging method have stimulated significant inspirations. The methods using space laser interferometers and using PTAs are becoming two important methods of detecting GWs. The PTAs and their sensitivity are addressed in Refs. [5, 7]. Interferometric space missions and their sensitivities will be addressed in the following section.

## 4. Interferometric Space Missions

In a Michelson interferometer, the wave front is split into two parts to go in two different paths and then the two wave fronts are recombined to interfere. For white light, Michelson had to match the two optical path lengths very precisely in order to have interference fringes. After laser was invented, the coherence length became longer. One could build unequal arm Michelson interferometer. An alternative configuration of the Michelson interferometer is the Mach-Zehnder Interferometer. Two-way Doppler tracking can be considered as an unequal arm Michelson interferometer; the local oscillator splits off a beam directing to the uplink spacecraft and the return beam from the spacecraft transponder interferes with the local oscillator. The phase (and frequency) of the beat is measured as a function of time. The Doppler response of a single link is given by (5). Using (5) the response of two-way Doppler tracking[69,70] is given by

$$y(t) = -(1 - \underline{\boldsymbol{k}} \cdot \underline{\boldsymbol{n}})\,\Psi(t) - 2(\underline{\boldsymbol{k}} \cdot \underline{\boldsymbol{n}})\,\Psi(t - (1 + \underline{\boldsymbol{k}} \cdot \underline{\boldsymbol{n}})L) + (1 + \underline{\boldsymbol{k}} \cdot \underline{\boldsymbol{n}})\,\Psi(t - 2L). \quad (8)$$

The three terms in (8) correspond, respectively, to the projected amplitude of the wave at the event of reception of the Doppler tracking signal at Earth, the transponding event at the spacecraft, and the emission event of the tracking signal from Earth.

Since the deviation of the speed of the electromagnetic wave from that of vacuum in plasma is inversely proportional to the square of the frequency, the time uncertainty due to solar wind or ionized gas in the microwave propagation is smaller in the Ka band (32 GHz) and X band (8.4 GHz) than S band (2.3 GHz). This is one of two motivations for Doppler tracking of Cassini spacecraft to use Ka band and X band for better noise performance. The other motivation is with shorter wavelength, the measurement precision increases. At optical frequency, the wavelength is more than 4-order smaller and the plasma effect is 8-order smaller. Therefore when better sensitivities in the optical path length measurement was needed in GW detection, the GW community started to use optical method. When sensitivity is increased, we need to suppress spurious noise below the aimed sensitivity level. This requires that (i) we reduce the acceleration noise and implement the drag-free technology; (ii) we reduce the laser noise as much as possible. The basic drag-free technology is now demonstrated by LISA Pathfinder.[11] For reducing laser noise, we need laser stabilization. The best way is to implementing absolute stabilization; e.g., to lock to an iodine molecular line. However, laser stabilization alone is not enough for the required strain sensitivity of the order of $10^{-21}$. To lessen the laser noise requirement, time delay interferometry (TDI) came to rescue.

For space laser-interferometric GW antenna, the arm lengths vary according to solar-system orbit dynamics. In order to attain the requisite sensitivity, laser frequency noise must be suppressed below the secondary noises such as the optical path noise, acceleration noise etc. For suppressing laser frequency noise, it’s necessary to use TDI in the analysis to match the optical path length of different beams closely. The better match of the optical path lengths are, the better cancellation of the laser frequency noise and the easier to achieve the requisite sensitivity. In case of exact match, the laser frequency noise is fully cancelled, as in the original Michelson interferometer.

The TDI was first used in the study of ASTROD mission concept.[23,25,26] In the deep-space interferometry, long distances are invariably involved. Due to long distances, laser light is attenuated to a great extent at the receiving spacecraft. To transfer the laser light back or to another spacecraft, amplification is needed. The procedure is to phase lock the local laser to the incoming weak laser light and to transmit the local laser light back or to another spacecraft. Liao *et al.* [29,30] have demonstrated the phase locking of a local oscillator with 2-pW laser light in laboratory. Dick *et al.*[83] have demonstrated phase locking to 40-fW incoming weak laser light. The power requirement feasibility for both e-LISA/NGO and ASTROD-GW is met with these developments. In the 1990s, Ni *et al.*[23,25,26] used the following two TDI configurations during the study of ASTROD interferometry and obtained numerically the path length differences using Newtonian dynamics.

These two TDI configurations are the unequal arm Michelson TDI configuration and the Sagnac TDI configuration for 3 spacecraft formation flight. The principle is to have two split laser beams to go to Path 1 and Path 2 and interfere at their end path. For unequal arm Michelson TDI configuration, one laser beam starts from spacecraft 1 (S/C1) directed to and received by spacecraft 2 (S/C2), and optical phase locking the local laser in S/C2; the phased locked laser beam is then directed to and received by S/C1, and optical phase locking another local laser in S/C1; and so on following Path 1 to return to S/C1:

Path 1: S/C1→S/C2→S/C1→S/C3→S/C1. (9)

The second laser beam starts from S/C1 also, but follows the Path 2 route:

Path 2: S/C1→S/C3→S/C1→S/C2→S/C1, (10)

to return to S/C1 and to interfere coherently with the first beam. If the two paths has exactly the same optical path length, the laser frequency noises cancel out; if the optical path length difference of the two paths are small, the laser frequency noises cancel to a large extent. In the Sagnac TDI configuration, the two paths are:

Path 1: S/C1→S/C2→S/C3→S/C1,
Path 2: S/C1→S/C3→S/C2→S/C1. (11)

Since then we have worked out the same things numerically for LISA,[84] eLISA/NGO,[85] LISA-type with $2 \times 10^6$ km arm length,[85] ASTROD-GW with no inclination,[86,87] and ASTROD-GW with inclination.[41]

Time delay interferometry has been worked out for LISA much more thoroughly on various aspects since 1999.[88,89] First-generation and second-generation TDIs are proposed. In the first generation TDIs, static situations are considered, while in the second generation TDIs, motions are compensated to certain degrees. The two configurations considered above are first generation TDI configurations in the sense of Armstrong, Estabrook and Tinto.[88,89] We will discuss numerical TDI more in Sec. 10. For many other aspects of TDI, we refer the readers to the excellent review [89].

In Table I we have compiled various interferometric space mission proposals for GW detection. Among the proposed science orbits, there are basically 3 categories – ASTROD-GW like, LISA like and OMEGA like.

(i) LISA-like (LAGOS-like)[14] science orbits: As in Fig. 1, the Earth-like solar orbits of the three spacecraft are appropriately inclined so that they form a nearly equilateral triangle formation having a tilt of ±60 degrees (in the figure, the tilt is 60 degrees) with respect to the ecliptic plane.[14] The formation rotates once per year clockwise or counterclockwise facing the Sun. Secs. 7 and 8 give more detailed orbit analysis. LISA,[9] eLISA,[21] ALIA[47] and TAIJI (ALIA-descope)[48] have this kind of LISA like science orbits. The ultimate configuration of Big Bang Observer[45] and DECIGO[44] has 12 spacecraft distributed in the Earth orbit in 3 groups separated by 120 degrees in orbit; 2 groups has 3 spacecraft each in a LISA-like triangular formation and the third group has 6 spacecraft with two LISA-like triangles forming a star configuration (Fig. 7). An alternate configuration is that each group has 4 spacecraft forming a nearly square configuration (also has a tilt of 60 degrees with respect to the ecliptic plane).

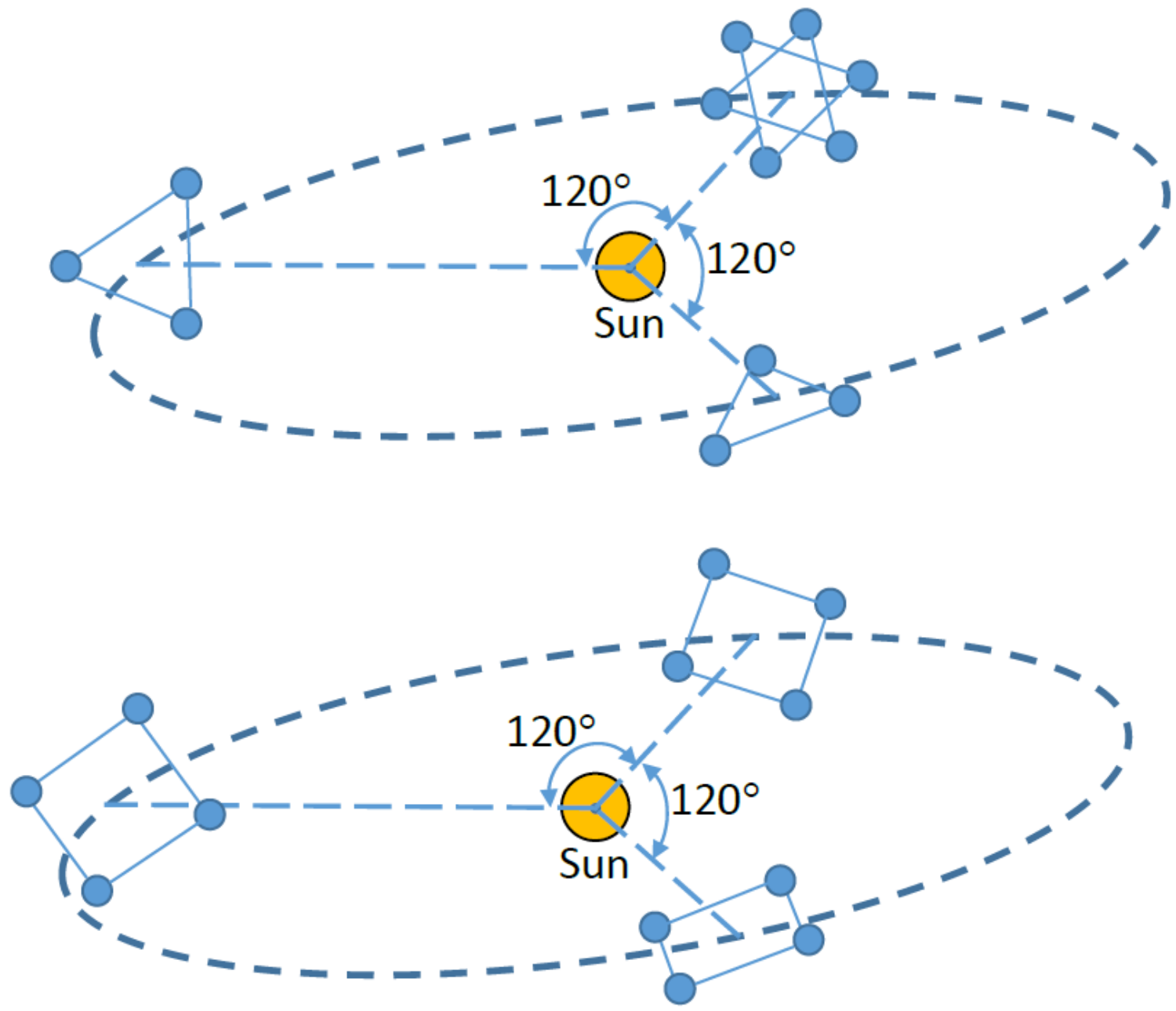

Fig. 7. Two schematic configurations of BBO and DECIGO in Earth-like solar orbits.

(ii) OMEGA-like science orbits: These orbits are Earth orbits away from (either inside or outside) Moon's orbit around the Earth. An example is the OMEGA mission orbit configuration. OMEGA mission proposed to NASA as a candidate MIDEX mission in 1998, and again as a mission-concept white paper in 2011. The OMEGA[54,55] mission consists of six identical spacecraft in a 600,000-km-high Earth orbit, two spacecraft at each vertex of a nearly equilateral triangle formation (Fig. 3). These orbits are stable, allowing for 3 years of planned science operations, as well as the possibility of an extended mission if desired. The arm length of the triangle formation is about 1 million km (1 Gm). The mission formation is outside of Moon's orbit.

There are 2 mission proposals -- GEOGRAWI[49]/gLISA[50,51] and GADFLI[52]using geostationary orbit formation. The 3 spacecraft of the formation are in the geostationary orbits forming a nearly equilateral triangle with arm length about 73,000 km.

TianQin is a GW mission proposal with 57,000 height orbit. The 3 spacecraft form a nearly equilateral triangle with arm length about 110,000 km with Earth-orbiting period 44 hours.[56]

The orbits and spacecraft configuration of all these missions are near ecliptic plane. There are times the Sun light comes along the line of sight of telescope links. Sunlight shields are required when the line of sight cross the Sun. A solution has been proposed from the OMEGA mission proposal[55] which could be used for other missions in this category.

(iii) ASTROD-GW like science orbits: The basic ASTROD-GW configuration consists of three spacecraft in the vicinity of the Sun-Earth Lagrange points L3, L4 and L5 respectively with near-circular orbits around the Sun, forming a nearly equilateral triangle as shown by Fig. 2 with the three arm lengths about $2.6\times10^8$ km (1.732 AU).[3,36-40] The dominant force on the spacecraft is from the Sun in the restricted three-body problem of Earth-Sun-spacecraft system. Since the Earth-Sun orbit is elliptical, the Lagrange points are not stationary in the Earth-Sun rotating frame. The motion of test particles at L3, L4 and L5 deviates from circular orbit by a fraction of O($e$) where $e$ (=0.0167) is the eccentricity of the Earth orbit around the Sun. However, the spacecraft can be in the halo orbit of the respective Lagrange points largely compensating the non-stationary motion of the Lagrange points to remain nearly circular orbits of the Sun. The circular orbits of spacecraft near the L3, L4 and L5 points are stable or virtually stable in 20 years (their orbits are also stable or quasi-stable with respect to their respective Lagrange points so that the deviations from circular orbit of their respective Lagrange point are of the order of O($e^2$) in AU ) and the deviation of the spacecraft triangle from an equilateral triangle is of order of O($e^2$) in arm length. For a non-precession planar formation, the angular resolution has antipodal ambiguity. To resolve this issue, we need to have precession orbit formation inclined with respect to the ecliptic. When the orbits of spacecraft have a small inclination $\lambda$ (in radians) with respect to the ecliptic plane the arm length variation is of the order of O($\lambda^2$). Therefore the added variation due to these two causes is of the order O($e^2$, $\lambda^2$). For these two causes to match (to O($10^{-4}$)), $\lambda$ should be of the order of O(1°). In Sec. 7, we review the inclined orbit analytically in the solar gravitational field and explain the angular resolution together with how to resolve the antipodal ambiguity. In Sec 8, we will use solar-system ephemeris to design and optimize the orbit configuration and will see that the perturbation from all planets except Earth is of the order of O($10^{-4}$). The influence of Earth is already taken into consideration since the L3, L4 and L5 points are effectively stable in 20 years. Hence, suitable inclined circular orbits could be our basic orbits to start with and the deviations from actual optimized orbit should be on the order of O($10^{-4}$).

For Super-ASTROD,[42] we could also place the 3 spacecraft with small inclination angle to Jovian solar orbit plane near Sun-Jupiter L3, L4 and L5 points with the other 1 or 2 spacecraft having large inclination(s).

For ASTROD-EM,[43] the 3 spacecraft will be placed in near Earth-Moon L3, L4 and L5 points. For the spacecraft dynamics, we have restricted 4-body (Earth, Moon, Sun, and the spacecraft whose gravitational filed can be neglected) problem to work out.

## 5. Frequency Sensitivity Spectrum

The space GW detectors are basically real-time free-mass detectors. As we have already discussed in [5] in general, there are two crucial issues in these proposed detectors: (i) to lower the disturbance effects and/or to model them for subtraction: drag-free to decrease the effects of surrounding disturbances, and appropriate modeling of the motion and the disturbances for subtraction to lower the residuals; (ii) to increase measurement sensitivity: microwave sensing, optical sensing, X-ray sensing, atom sensing, molecule sensing and timing…. Associated with these two issues, there are two basic noises – the acceleration noise and the metrology noise. For laser-optic missions, the metrology noise is the laser metrology noise. The planned upper limits of these two kinds of basic noise for GW mission proposals are listed in the last two columns of Table 1. In space GW detection, the basic noise model is the LISA/eLISA noise model. Due to more stringent technological requirements, Big Bang Observer and DECIGO belong to second generation space detector proposals. All others in Table 1 are first generation space detector proposals. In Figs. 4-6, we plot the sensitivity curves of three typical first generation space detectors (LISA/eLISA, ASTROD-GW and OMEGA) and two second generation space detectors (Big Bang Observer and DECIGO). In the first generation category, for missions with arm length shorter than LISA, the planned strain upper limits are smaller than that of LISA in the higher frequency part; for missions with arm length longer than LISA, the planned strain upper limits are smaller than that of LISA in the lower frequency part.

As shown in Fig. 4, typical frequency sensitivity spectrum of strain psd amplitude for space GW detection consists of 3 regions, the acceleration/vibration noise dominated region, the shot noise (flat for current space detector projects like LISA in strain psd) dominated region, if any, and the antenna response restricted region. The lower frequency region for the detector sensitivity is dominated by vibration, acceleration noise or gravity-gradient noise. The higher frequency part of the detector sensitivity is restricted by antenna response (or storage time). In a power-limited design, sometimes there is a middle flat region in which the sensitivity is limited by the photon shot noise.[9,23,40]

The shot noise sensitivity limit in the strain for GW detection is inversely proportional to $P^{1/2}L$ with $P$ the received power and $L$ the distance or arm length. Since $P$ is inversely proportional to $L^2$ and $P^{1/2}L$ is constant, this sensitivity limit is independent of the distance. For 1-2 W emitting power, the limit is around $10^{-21}$ $\mathrm{Hz}^{-1/2}$. As noted in the LISA study,[9] making the arms longer shifts the time-integrated sensitivity curve to lower frequencies while leaving the bottom of the curve at the same level. Hence, ASTROD-GW with longer arm length has better sensitivity at lower frequency. e-LISA, ALIA, TAIJI (ALIA-descope), and GW interferometers in Earth orbit have shorter arms and therefore have better sensitivities at higher frequency.

In Fig.'s 4-6, we plot sensitivity curves for LISA, e-LISA and ASTROD-GW for the low-frequency GW band. In the Mock LISA Data Challenge (MLDC) program, the consensus goal for the LISA instrumental noise density amplitude ${}^{(\mathrm{MLDC})}S_{\mathrm{Ln}}^{1/2}(f)$ is

$$^{(\mathrm{MLDC})}S_{\mathrm{Ln}}^{1/2}(f) = (1/L_{\mathrm{L}}) \times \{[(1 + 0.5\,(f / f_{\mathrm{L}})^2\,)] \times S_{\mathrm{Lp}} + [1 + (10^{-4}\ \mathrm{Hz} / f)^2]\,(4S_{\mathrm{a}}/(2\pi f)^4)\}^{1/2}\ \mathrm{Hz}^{-1/2}, \quad (12\mathrm{a})$$

where $L_{\mathrm{L}} = 5 \times 10^9$ m is the LISA arm length, $f_{\mathrm{L}} = c / (2\pi L_{\mathrm{L}})$ is the LISA arm transfer frequency, $S_{\mathrm{Lp}} = 4 \times 10^{-22}$ m$^2$ Hz$^{-1}$ is the LISA (white) position noise (power) level due to photon shot noise, and $S_{\mathrm{a}} = 9 \times 10^{-30}$ m$^2$ s$^{-4}$ Hz$^{-1}$ is the LISA white acceleration noise (power) level.[90] Note that (12a) contains the "reddening" factor $[1 + (10^{-4}\ \mathrm{Hz} / f)^2]$ in the acceleration noise term.

In 2003, Bender[91] looked into the possible LISA sensitivity below 100 μHz. From a careful analysis of noises of test mass and capacitive sensing, Bender suggested a specific sensitivity goal at frequencies down to 3 μHz which contained a milder (than MLDC) "reddening factor". For frequency between 10 μHz to 100 μHz, he suggested to put in the "reddening factor" $[(10^{-4}\ \mathrm{Hz} / f)^{1/2}]$ and for frequency between 3 μHz to 10 μHz, the "reddening factor" $[3.16 \times (10^{-5}\ \mathrm{Hz} / f)]$. To drop this "reddening factor" might be difficult. However, with monitoring the gap of capacitive sensing and the positions of major mass distribution, the factor may be alleviated to certain extent. To completed drop the factor or to go beyond, one may need to go to optical sensing and optical feedback control.[24,27,28,92-94] If we drop the "reddening factor", the enhanced LISA instrumental noise density amplitude $^{(\mathrm{Enhanced})}S_{\mathrm{Ln}}^{1/2}(f)$ becomes

$$^{(\mathrm{Enhanced})}S_{\mathrm{Ln}}^{1/2}(f) = (1/L_{\mathrm{L}}) \times \{[(1 + 0.5\,(f / f_{\mathrm{L}})^2\,)] \times S_{\mathrm{Lp}} + [4S_{\mathrm{a}}/(2\pi f)^4]\}^{1/2}\ \mathrm{Hz}^{-1/2}. \quad (12\mathrm{b})$$

After NASA's withdrawal from ESA-NASA collaboration of LISA in 2011, the European eLISA/NGO (NGO: New Gravitational-wave Observatory) for space detection of GWs emerged. The orbit configuration is the same as LISA, but with arm length shrunk 5 times to one million kilometers, the orbits slowly drifting away from the Earth and the nominal mission duration 2 years (extendable to 5 years) to save weight, fuel and costs. The three spacecraft will consist of one "mother" and two simpler "daughters," with interferometric measurements along only two arms with the "mother" at the vertex.[21] The eLISA/NGO strain noise power-spectral-density goal is also shown in Fig. 4. For the lower frequency part of the power spectrum of eLISA/NGO, we choose to use the same acceleration noise with reddening factor (solid line) and without reddening factor (dashed line) as those of LISA to obtain the eLISA/NGO strain noise for easy comparison.

The eLISA arm length $L_{e\mathrm{L}}$ is 5 times shorter. Its instrumental noise density amplitude $^{(\mathrm{MLDC})}S_{e\mathrm{Ln}}^{1/2}(f)$ is

$$^{(\mathrm{MLDC})}S_{e\mathrm{Ln}}^{1/2}(f) = (1/L_{e\mathrm{L}}) \times \{[(1 + 0.5\,(f / f_{e\mathrm{L}})^2\,)] \times S_{e\mathrm{Lp}} + [1+(10^{-4}\ \mathrm{Hz} / f)^2](4S_{\mathrm{a}}/(2\pi f)^4)\}^{1/2}\ \mathrm{Hz}^{-1/2}, \quad (13\mathrm{a})$$

where $L_{e\mathrm{L}} = 10^9$ m is the eLISA arm length, $f_{e\mathrm{L}} = \mathrm{c} / (2\pi L_{e\mathrm{L}})$ is the *e*LISA arm transfer frequency, $S_{e\mathrm{Lp}} = 1 \times 10^{-22}$ m$^2$ Hz$^{-1}$ is the eLISA (white) position noise level due to photon shot noise assuming that the telescope diameter is 25 cm (compared with 40 cm for that of LISA) and that the laser power is the same as LISA. With these assumptions, the eLISA position noise amplitude would be 10 pm/Hz$^{1/2}$ listed in parentheses in the eLISA entry, comparable to 12 pm/Hz$^{1/2}$ used in Ref. [95]. The corresponding enhanced eLISA instrumental noise density amplitude ${}^{(\mathrm{Enhanced})}S_{e\mathrm{Ln}}^{1/2}(f)$ is

$$^{(\mathrm{MLDC})}S_{e\mathrm{Ln}}^{1/2}(f) = (1/L_{e\mathrm{L}}) \times \{[(1 + 0.5\,(f / f_{e\mathrm{L}})^2)] \times S_{e\mathrm{Lp}} + (4S_{\mathrm{a}}/(2\pi f)^4)\}^{1/2}\ \mathrm{Hz}^{-1/2}. \quad (13b)$$

For ASTROD-GW, our goal on the instrumental strain noise density amplitude is

$$S_{\mathrm{An}}^{1/2}(f) = (1/L_{\mathrm{A}}) \times \{[(1 + 0.5\,(f / f_{\mathrm{A}})^2)] \times S_{\mathrm{Ap}} + [4S_{\mathrm{a}}/(2\pi f)^4]\}^{1/2}\ \mathrm{Hz}^{-1/2}, \quad (14)$$

over the frequency range of 100 nHz < $f$ < 1 Hz. Here $L_{\mathrm{A}} = 260 \times 10^9$ m is the ASTROD-GW arm length, $f_{\mathrm{A}} = c\,/\,(2\pi L_{\mathrm{A}})$ is the ASTROD-GW arm transfer frequency, $S_{\mathrm{a}} = 9 \times 10^{-30}\ \mathrm{m}^2\ \mathrm{s}^{-4}\ \mathrm{Hz}^{-1}$ is the white acceleration noise level (the same as that for LISA), and $S_{\mathrm{Ap}} = 10816 \times 10^{-22}\ \mathrm{m}^2\ \mathrm{Hz}^{-1}$ is the (white) position noise level due to laser shot noise which is 2704 ($=52^2$) times that for LISA.[3,36-40] The corresponding noise curve for the ASTROD-GW instrumental noise density amplitude ${}^{(\mathrm{MLDC})}S_{\mathrm{An}}^{1/2}(f)$ with the same "reddening" factor as specified in MLDC program is

$$^{(\mathrm{MLDC})}S_{\mathrm{An}}^{1/2}(f) = (1/L_{\mathrm{A}}) \times \{[(1 + 0.5\,(f / f_{\mathrm{A}})^2)] \times S_{\mathrm{Ap}} + [1 + (10^{-4} / f)^2]\,(4S_{\mathrm{a}}/(2\pi f)^4)\}^{1/2}\ \mathrm{Hz}^{-1/2}, \quad (14a)$$

over the frequency range of 100 nHz < $f$ < 1 Hz. The sensitivity curves from the six formulas (12a,b) to (14a) are shown in Fig. 4. The corresponding sensitivity curves in terms of $h_c(f)$ and $\Omega_{\mathrm{gw}}(f)$ are shown in Fig. 5 and Fig. 6 respectively. The ones with reddening factor are shown with dashed line in the lower frequency part.

With the same laser power as that of LISA, the ASTROD-GW sensitivity would be shifted to lower frequency by a factor up to 52 if other frequency-dependent requirements can be shifted and met. The sensitivity curve would then be shifted toward lower frequency as a whole. Since the main constraints on the lower frequency part of the sensitivity is from the accelerometer noise, this translational shift depends on whether the accelerometer noise requirement for ASTROD-GW could be lowered (more stringent) from that of LISA requirement at a particular frequency. Since ASTROD is in a time frame later than LISA, if the absolute metrological accelerometer/inertial sensor could be developed, there is a potential to go toward this requirement. However, *to be simple*, we have taken a conservative stand and assume that *the LISA accelerometer noise goal and all other local requirements are taken as they are in the above equations and in the plotting of sensitivity curves in Fig.'s 4-6*. Since the strain sensitivity is mainly the accelerometer noise divided by arm length at low frequency, at a particular low frequency limited by accelerometer noise, the strain sensitivity for ASTROD-GW is 52 times lower than LISA (or 260 times lower than eLISA) due to longer arm length whether we take (12a) [or (13a)] and (14a) to compare or (12b) [or (13b)] and (14) to compare. With better lower-frequency resolution, the confusion limit of Galactic compact binary background for ASTROD-GW would be somewhat lower than that for LISA. The confusion limit for eLISA would be somewhat higher than that for LISA. In Fig.'s 4-6, the confusion limit curves are for LISA. ASTROD-GW will complement LISA and PTAs in exploring single events and backgrounds of MBH-MBH binary GWs in the important frequency range 100 nHz - 1 mHz to study black hole co-evolution with galaxies, dark energy and other issues (Sec. 6).

OMEGA has one million km arms just as eLISA. The sensitivity goal of OMEGA is: (i) The acceleration noise psd is the same as LISA and eLISA; (ii) the (white) position noise amplitude is fourfold lower than LISA and twofold lower than eLISA. The sensitivity curve of OMEGA plotted on Fig. 4 is from Ref. [55] with corresponding curves shown on Fig. 5 and Fig. 6. The lower frequency part and the flat part are close to eLISA while the antenna-response-limited part is slightly better. The small difference as compared to the Table may because of OMEGA has three pairs of S/C with one more link for interferometry or just because of different sources of drawing.

For GW mission proposals listed in Table 1 with formations inside the lunar orbit around the Earth, the acceleration noise requirements are about the same level or slightly more stringent than OMEGA and eLISA while the requirement on the position noise amplitude is lower because of more power received. The goal sensitivity curves in the higher frequency part is slightly better for the two mission proposals in geostationary orbits, gLISA/GEOGRAWI and GADFLI. TIANQIN with 0.11 Gm arm length aims at first and sure detection of a GW source in space, the required sensitivity on $S_a^{1/2}$ and $S_p^{1/2}$ are $S_a^{1/2} = 1 \times 10^{-15}$ m s$^{-2}$ Hz$^{-1/2}$ and $S_p^{1/2} = 1$ pm Hz$^{-1/2}$ at 6 mHz.

ALIA in solar orbit as a LISA follow-on aims at better sensitivity at frequency above 1 mHz. It has arm length of 0.5 Gm (05 million km) – ten times shorter than LISA and two times shorter than eLISA. The acceleration noise requirement is tenfold more stringent than LISA, i.e. $S_a^{1/2} = 0.3 \times 10^{-15}$ m s$^{-2}$ Hz$^{-1/2}$. The position noise amplitude requirement is 30 times more stringent than LISA, i.e. $S_p^{1/2} = 0.6 \times 10^{-15}$ pm Hz$^{-1/2}$. TAIJI (ALIA-descope) has arm length of 3 Gm and aims at a detection of intermediate black hole coalescence in addition to other scientific goals common to most space mission proposals. Its sensitivity is relaxed from ALIA to $S_a^{1/2} = 3 \times 10^{-15}$ m s$^{-2}$ Hz$^{-1/2}$ (the same as LISA) and $S_p^{1/2} = 5\text{-}8$ pm Hz$^{-1/2}$.

The 3 spacecraft of ASTROD-EM and of LAGRANGE will be located near L3, L4 and L6 Lagrange points of Earth-Moon system respectively. Due to the inclination of the Moon-Earth orbit plane to the ecliptic, the spacecraft formation plane will not intersect the Sun. Hence unlike other missions in Earth orbit, the Sun light will not come along the line of sight of telescope links. Sunlight shields are not required. The spacecraft dynamics is a restricted 4-body (Earth, Moon, Sun and the spacecraft) problem which we are still working on.[43] The acceleration noise and the laser metrology noise requirements are listed in Table 1.

BBO and DECIGO have similar goals of detecting primordial GWs. BBO has a delay line implementation. DECIGO uses a Fabry-Perot implementation. The acceleration noise $S_a^{1/2}$ and the laser metrology noise $S_p^{1/2}$ requirements of DECIGO are $S_a^{1/2} = 3 \times 10^{-17}$ m s$^{-2}$ Hz$^{-1/2}$ and $S_p^{1/2} = 1.4 \times 10^{-5}$ pm Hz$^{-1/2}$ respectively. The strain sensitivity curve of a single DECIGO interferometer as shown in Fig. 5 is from [96]. BBO has a similar single-interferometer sensitivity curve. One-sigma, power-law integrated sensitivity curve for BBO (BBO-corr) as shown in Fig. 5 is obtained by Thrane and Romano [97]. That of DECIGO is similar. We also put in the plot their LISA autocorrelation measurement sensitivity curve (LISA-corr) in a single detector assuming perfect subtraction of instrumental noise and/or any unwanted astrophysical foreground.[97] The minimum autocorrelation sensitivity using the same method for ASTROD-GW is also estimated and plotted in Fig. 5; this would also be the level that 6 S/C ASTROD-GW[40] (6 S/C ASTROD-GW-corr) could reach. All of the corresponding curves are plotted in Fig. 4 and Fig. 6. *Considering the sensitivity requirements or arm length involved, DECIGO, BBO and Super-ASTROD belongs to the second-generation space interferometers.* For the sensitivity of Super-ASTROD, we assume $S_a^{1/2} = 3 \times 10^{-15}$ m s$^{-2}$ Hz$^{-1/2}$ (the same as LISA) and $S_p^{1/2} = 5000$ pm Hz$^{-1/2}$.

*Atom Interferometry.* The development in atom interferometry is fast and promising. It already contributes to precision measurement and fundamental physics. A proposal using atom interferometry to detect GWs has been raised at Stanford University as an alternate method to LISA on the LISA bandwidth.[98,99] Issues have arisen on its realization of LISA sensitivity for this proposal.[100,101] In Observatoire de Paris, SYRTE has started the first stage of its project -- MIGA (Matter-wave laser Interferometric Gravitation Antenna)[102] of building a 300-meter long optical cavity to interrogate atom interferometers at the underground laboratory LSBB (Laboratoire Souterrain à Bas Bruit) in Rustrel. In the second stage of the project (2018-2023), MIGA will be dedicated to science runs and data analyses in order to probe the spatio-temporal structure of the local field of the LSBB region. In the meantime, MIGA will assess future potential applications of atom interferometry to GW detection in the middle frequency band (0.1-10 Hz).

## 6. Scientific Goals

In this section, we review and summarize the scientific goals for space GW mission proposals and projects.[3,9,21,39,40,95] More studies on the scientific goals and data analysis in the next few years will be worthy for the preparation of space GW missions.

### 6.1. Massive Black Holes (MBHs) and their co-evolution with galaxies

Relations have been discovered between the MBH mass and the bulge mass of host galaxy, and between the MBH mass and the velocity-dispersion of host galaxy. These relations indicate that the central MBHs are linked to the evolution of galactic structure. Observational evidence indicate that MBHs reside in most local galaxies. Newly fueled quasar may come from the gas-rich major merger of two massive galaxies. GW observation in the low frequency band (100 nHz - 100 mHz) by space interferometers and very low frequency band (300 pHz-100 nHz) by PTAs will be a major tool to study the co-evolution of galaxy with BHs.

The standard theory of massive black hole formation is the merger-tree theory with various Massive Black Hole Binary (MBHB) inspirals acting. The GWs from these MBHB inspirals can be detected and explored to cosmological distances using space GW detectors and PTAs depending on the masses of MBHBs. Although there are different merger-tree models and models with BH seeds, they all give significant detection rates for space GW detectors and Pulsar Timing Arrays (PTAs),[7,103-105] NGO/eLISA[21] and ASTROD-GW.[40] PTAs are most sensitive in the frequency range 300 pHz-100 nHz, NGO/eLISA space GW detector is most sensitive in the frequency range 2 mHz – 0.1 Hz, while ASTROD-GW is most sensitive in the frequency range 100 nHz - 2 mHz (Fig.'s 4-6). NGO/eLISA and ASTROD-GW will be able to directly observe how massive black holes form, grow, and interact over the entire history of galaxy formation. ASTROD-GW will detect stochastic GW background from MBH binary mergers in the frequency range 100 nHz to 100 μHz. These observations are significant and important to the study of co-evolution of galaxies with MBHs. The expected rate of MBHB sources is 10 $yr^{-1}$ to 100 $yr^{-1}$ for NGO/eLISA and 10 $yr^{-1}$ to 1000 $yr^{-1}$ for LISA.[21] For ASTROD-GW, we are expecting similar number of sources but with better angular resolution (Sec. 7.3).[40]

A sample of MBHB merger sources are drawn on Fig.'s 4-6. The black lines show the inspiral, coalescence and oscillation phases of GW emission from various equal-mass black-hole binary mergers in circular orbits at various redshift: solid line, $z$ = 1; dashed line, $z$ = 5; long-dashed line $z$ = 20. The $10^6$ M⊙-$10^6$ M⊙ MBHB merger at $z$ = 1, $10^5$ M⊙-$10^5$ M⊙ MBHB merger at $z$ = 20, and $10^4$ M⊙-$10^4$ M⊙ MBHB merger at $z$ =5 are from Schutz [106] for Fig. 4; others by scaling; the corresponding curves in Fig. 5 and Fig. 6 by transformation Equation (1). MBHB merger events have large signal to noise ratio for space detectors. Some of these events with equal mass (from $10^2$-$10^{10}$ M⊙) and circular orbit are shown in Fig.'s 4-6. They are all candidates for space-borne detectors. Some could be in earlier phases for future ground-based detectors.

With the detection of MBHB merger events and background, the properties and distribution of MBHs could be deduced and underlying population models could be tested.

PTAs have been collecting data for decades for detection of stochastic GW background from MBHB mergers. In modeling the MBHB stochastic GW background spectra, various authors obtained the following frequency dependence:

$$h_c(f) = A_{\rm yr}\, [f/(1\ {\rm yr}^{-1})]^{\alpha}, \qquad (15)$$

with $\alpha = -\,(2/3)$.[7] PTAs have improved greatly on the sensitivity for GW detection recently.[107-109] They have put upper limits on the isotropic stochastic background assuming the frequency dependence (15) with $\alpha = -\,(2/3)$ as follows: from EPTA (European PTA), $A_{\rm yr} < 3 \times 10^{-15}$; from PPTA (Parks PTA), $A_{\rm yr} < 1 \times 10^{-15}$ and NANOGrav (North American Nanohertz Observatory for Gravitational Waves), $A_{\rm yr} < 1.5 \times 10^{-15}$. The three experiments form a robust upper limit of $1 \times 10^{-15}$ on $A_{\rm yr}$ at 95 % confidence level ruling out most models of supermassive black hole formation. The limit is shown as constraint on the Supermassive Black Hole Binary GW Background (SBHB-GWB) in Fig.s 2-4 of [5] as solid line in the frequency range $10^{-9}$- $10^{-7}$ Hz. The GW energy released from co-evolution with galaxies must go somewhere. More energy of GWs might be emitted with higher frequency in the hierarchy of supermassive black hole formation. Hence we have extrapolated this constraint linearly (instead with a knee around $f \sim 100$ nHz in most existed models) with dotted line to 10 μHz with some confidence in our review [5]. We adopt the same thing in Fig.'s 4-6 here. Constraints with other $\alpha$ values have similar order of magnitudes.

## 6.2. Extreme Mass Ratio Inspirals (EMRIs)

EMRIs are GW sources for space GW detectors. The NGO/eLISA sensitive range for central MBH masses is $10^4$-$10^7$ $M_\odot$. The expected number of NGO/eLISA detections over two years is 10 to 20;[21] for LISA, a few tens;[21] for ASTROD-GW, similar or more with sensitivity toward larger central BH's and with better angular resolution (Sec. 7.3).[40]

## 6.3. Testing Relativistic Gravity

An important scientific goal of LISA[21,9] and NGO/eLISA[21,95] is to test general relativity and to study black hole physics with precision in strong gravity. With better precision in 100 nHz-1 mHz frequency range, ASTROD is going to push this goal further in many aspects. These include testing strong-field gravity, precision probing of Kerr spacetime and measuring/constraining the mass of graviton. Some considerations have been given in [110, 111]. Lower frequency sensitivity is significant in improving the precision of various tests.[110,111] Further studies in these respects would be of great value.

## 6.4. Dark energy and cosmology

In the dark energy issue,[112] it is important to determine the value of $w$ in the equation of state of dark energy,

$$w = p / \rho, \tag{16}$$

as a function of different epochs where $p$ is the pressure and $\rho$ the density of dark energy. For cosmological constant as dark energy, $w = -1$. From cosmological observations, our universe is close to being flat. In a flat Friedman-Lemaître-Robertson-Walker (FLRW) universe, the luminosity distance is given by

$$d_L(z) = (1+z)(H_0)^{-1} \int_0^z dz' \left[\Omega_m (1+z')^3 + \Omega_{\rm DE} (1+z')^{3(1+w)}\right]^{-\frac{1}{2}}, \tag{17}$$

where $H_0$ is Hubble constant, $\Omega_{DE}$ is the present dark energy density parameter, and the equation of state of the dark energy $w$ is assumed to be constant. In the case of non-constant $w$ and non-flat FLRW universe, similar but more complicated expression can be derived. Here we show (17) for illustrative purpose. From the observed relation of luminosity distance vs. redshift $z$, the parameter $w$ of the equation of state as a function of redshift $z$ can be solved for and compared with various cosmological models. Dark energy cosmological models can be tested this way. Luminosity distance from supernova observations and from gamma ray burst observations vs. redshift observations are the focus for the current dark energy probes.

Space GW detectors observing MBHB inspirals and EMRIs are good probes to determine the luminosity distances. With the redshift of the source determined by the electromagnetic observations of associated galaxies or cluster of galaxies, these space

GW detectors are also dark energy probes. In the merging of MBHs during the galaxy co-evolution processes, gravitational waveforms generated give precise, gravitationally calibrated luminosity distances to high redshift. The inspiral signals of these binaries can serve as standard candles/sirens.[113,114] With better angular resolution (Sec. 7.3), ASTROD-GW will have better chance to identify the associated electromagnetic redshift and therefore will be better for the determination of the dark energy equation of state.[39,40]

### 6.5. Compact binaries

Space GW detectors are also sensitive to the GWs from Galactic compact binaries.[9,21] These detectors will be able to survey compact stellar-mass binaries and study the structure of the Galaxy. NGO/eLISA will detect about 3000 double white dwarf binaries individually with most in the GW frequency band 3-6 mHz (orbit period about 300-600 s); for LISA, about 10,000 double white dwarf binaries.[9,21] These sources constitute the population which has been proposed as progenitors of normal type Ia and peculiar supernovae. For a review on the electromagnetic counterparts of GW mergers of compact objects, see, e.g., Ref. 115. At the frequency band 3-6 mHz, NGO/eLISA is more sensitive than ASTROD-GW (Fig. 4). Since NGO/eLISA will be flying first these GW signals will serve as a calibration for ASTROD-GW in addition to the verification binaries. The 8 verification binaries selected by NGO/eLISA are shown on Fig. 4 as red squares with 2-year integration time (from Ref. 21, p. 14, Figure 2.6).

At GW frequencies below a few mHz, millions of ultra-compact binaries will form a detectable foreground for NGO/eLISA and ASTROD-GW. At these frequencies, ASTROD-GW is more sensitive than NGO/eLISA (Fig. 4). More sources will be resolved individually and ASTROD-GW can improve on the observational results of NGO/eLISA.

### 6.6. Relic GWs

For direct detection of primordial (inflationary, relic) GWs in space, one may go to frequencies lower or higher than LISA bandwidth,[3,116] where there are potentially less foreground astrophysical sources[117] to mask detection. DECIGO[44] and Big Bang Observer[45] look for GWs in the higher frequency range while ASTROD-GW[3,116] looks for GWs in the lower frequency range. Their instrument sensitivity goals all reach $10^{-17}$ in terms of critical density. The main issue is the level of foreground and whether foreground could be separated.

The straight line in the bottom left corner of Figure 4 corresponds to $\Omega_{gw} = 10^{-15}$ CMB (Cosmic Polarization Background) upper limit (See, e.g. [5]) of inflationary GW background. For ASTROD-GW, when a 6-S/C formation is used for correlated detection of stochastic GWs, the sensitivity can reach this region. However, the anticipated upper limit of MBH-MBH GW background is above the 3-S/C ASTROD-GW sensitivity. If this background is detected, then the detectability of inflationary gravitational wave of the strength $\Omega_{gw} = 10^{-16}$-$10^{-17}$ from 6-S/C formation in the ASTROD-GW frequency region depends on whether this MBH-MBH GW 'foreground' could be separated due to different frequency dependence or other signatures.[40]

Other potentially possible GW sources in the relevant frequency band, e.g., cosmic strings, should also be studied. See Ref. [95] for cosmic strings and some other sources.

## 7. Basic Orbit Configuration, Angular Resolution and Muli-Formation Configurations

In this section, we review and summarize the basic LISA-like and ASTRO-GW configurations, their angular resolutions and multi-formation configurations. These basic configurations can be used for starting numerical design and numerical orbit optimization for missions in these two categories.

### 7.1. Basic LISA-like orbit configuration

As in Fig. 1, the center of mass of the basic LISA-like configuration[9,14,85,118-123] follows a circular orbit of radius $R$ (= 1 AU) around the Sun. Since the distance (arm length) $L$ between the spacecraft is much smaller than the circular orbit radius 1 AU, we could treat the spacecraft orbits as perturbed orbits from the circular orbit. The equations for the perturbed orbit are known as Euler-Hill equations, Hill equations, or Clohessy and Wiltshire equations. Hill used these equations for researches in the lunar theory in the $19^{th}$ century.[124] Clohessy and Wiltshire[125] derived and used these equations for designing terminal guiding system for satellite rendezvous in 1960 after the space era began at 1957. Clohessy and Wiltshire used a frame – called $\mathcal{CW}$ frame with its origin on the circular reference (center of configuration) orbit and with the frame rotating with angular velocity $\Omega$ the same as that of reference orbit rotation. For the perturbed orbit to keep the same distance to the origin and to remain stationary in the $\mathcal{CW}$ frame, it is clear by calculation of the difference of the perturbed orbit and fiducial orbit that the eccentricity $e$ and the inclination $i$ with respect to the ecliptic need to be

$$e = 3^{-1/2}L \,/\, (2R);\ i = \alpha \equiv L/(2R)\ ; \tag{18}$$

to first order in the perturbation or to $O(L/(2R))$ [= $O(\alpha)$]. One way to form a nearly triangular configuration with side or arm length $L(1 + O(\alpha))$ is to require the orbit nodes be separated by 120°, and to choose the true anomalies and arguments of perihelion such that each spacecraft at its aphelion is also at its maximum height above (north of) the ecliptic ($1^{st}$ configuration); the other way is with aphelion at its minimum height below (south of) the ecliptic ($2^{nd}$ configuration).[9] With these choices, the mission configuration plane is at 60° from the ecliptic with the intersection to ecliptic tangential to the fiducial (center of configuration) circular orbit. For square configuration, just require the orbit nodes to be separated by 90°; one can similarly construct any regular polygon configuration or any planar configuration. The 1st configuration rotates clockwise; the 2nd rotates counterclockwise. Thus, one reaches the conclusion -- *in the $\mathcal{CW}$ frame there are just two planes which make angles of* ±60° *with the reference orbit plane, in which* *spacecraft (test particles) obeying the CW equations and perform rigid rotations about the origin with angular velocity* $-\Omega$.

We follow Dhurandhar *et al*.,[123] and Wang and Ni[85] to write down the equation for the basic orbits of the three spacecraft for the LISA-like configurations. First, the equation of an elliptical orbit in the general $X$-$Y$ plane is given by

$$X = R(\cos\psi + e),\ Y = R(1 - e^2)^{1/2}\sin\psi, \tag{19}$$

where $R$ is the semi-major axis of the ellipse, $e$ the eccentricity and $\psi$ the eccentricity anomaly.

Define $\alpha$ to be the ratio of the planned arm length $L$ of the orbit configuration to twice radius $R$ (1 AU) of the mean Earth orbit around the Sun, i.e., $\alpha = L/(2R)$. Choose the initial time $t_0$ to be a specific epoch in the Julian calendar and work in the Heliocentric Coordinate System ($X$, $Y$, $Z$). $X$-axis is in the direction of vernal equinox. A set of elliptical S/C orbits can be defined as

$$\begin{aligned} X_{\mathrm{f}} &= R(\cos\psi_{\mathrm{f}} + e)\cos\varepsilon, \\ Y_{\mathrm{f}} &= R(1-e^2)^{1/2}\sin\psi_{\mathrm{f}}, \\ Z_{\mathrm{f}} &= R(\cos\psi_{\mathrm{f}} + e)\sin\varepsilon. \end{aligned} \tag{20}$$

Here $R$ = 1 AU; $e$ = 0.001925; $\varepsilon$ = 0.00333. The eccentric anomaly $\psi_{\mathrm{f}}$ is related to the mean anomaly $\Omega\,(t-t_0)$ by

$$\psi_{\mathrm{f}} + e\sin\psi_{\mathrm{f}} = \Omega\,(t\text{-}t_0). \tag{21}$$

Here $\Omega$ is defined as $2\pi$/(one sidereal year). The eccentric anomaly $\psi_{\mathrm{f}}$ can be solved by numerical iteration. Define $\psi_k$ to be implicitly given by

$$\psi_k + e\sin\psi_k = \Omega\,(t - t_0) - 120^\circ(k-1),\ \text{for}\ k = 1, 2, 3. \tag{22}$$

Define $X_{\mathrm{f}k}$, $Y_{\mathrm{f}k}$, $Z_{\mathrm{f}k}$, ($k$ = 1, 2, 3) to be

$$\begin{aligned} X_{\mathrm{f}k} &= R(\cos\psi_k + e)\cos\varepsilon, \\ Y_{\mathrm{f}k} &= R(1-e^2)^{1/2}\sin\psi_k, \\ Z_{\mathrm{f}k} &= R(\cos\psi_k + e)\sin\varepsilon. \end{aligned} \tag{23}$$

Define $\varphi_0 \equiv \psi_{\mathrm{E}} - 10^{\circ}$ with $\psi_{\mathrm{E}}$ is the position angle of Earth with respect to the $X$-axis at $t_0$. Define $X_{\mathrm{f}(k)}$, $Y_{\mathrm{f}(k)}$, $Z_{\mathrm{f}(k)}$, ($k$ = 1, 2, 3), i.e. $X_{\mathrm{f}(1)}$, $Y_{\mathrm{f}(1)}$, $Z_{\mathrm{f}(1)}$; $X_{\mathrm{f}(2)}$, $Y_{\mathrm{f}(2)}$, $Z_{\mathrm{f}(2)}$; $X_{\mathrm{f}(3)}$, $Y_{\mathrm{f}(3)}$, $Z_{\mathrm{f}(3)}$ to be

$$X_{\mathrm{f}(k)} = X_{\mathrm{f}k} \cos[120°(k\text{-}1)+\varphi_0] - Y_{\mathrm{f}k} \sin[120°(k\text{-}1)+\varphi_0],$$
$$Y_{\mathrm{f}(k)} = X_{\mathrm{f}k} \sin[120°(k\text{-}1)+\varphi_0] + Y_{\mathrm{f}k} \cos[120°(k\text{-}1)+\varphi_0], \quad (24)$$
$$Z_{\mathrm{f}(k)} = Z_{\mathrm{f}k}.$$

The basic orbits of the three S/C are (for one-body central problem) are

$$\mathbf{R}_{\mathrm{S/C1}} = (X_{\mathrm{f}(1)}, Y_{\mathrm{f}(1)}, Z_{\mathrm{f}(1)}),$$
$$\mathbf{R}_{\mathrm{S/C2}} = (X_{\mathrm{f}(2)}, Y_{\mathrm{f}(2)}, Z_{\mathrm{f}(2)}), \quad (25)$$
$$\mathbf{R}_{\mathrm{S/C3}} = (X_{\mathrm{f}(3)}, Y_{\mathrm{f}(3)}, Z_{\mathrm{f}(3)}).$$

The initial positions can be obtained by choosing $t = t_0$ and initial velocities by calculating the derivatives with respect to time at $t = t_0$. As an example, if we choose $t_0$ = JD2459215.5 (2021-Jan-1st 00:00:00), the initial conditions (states) of three spacecraft of eLISA/NGO in J2000.0 solar-system-barycentric Earth mean equator and equinox coordinates are calculated and tabulated in the third column of Table 2 (from Table 2 of Ref. [85]). From these initial conditions, one could start to design and optimize the orbit configuration numerically using planetary and lunar ephemeris as in Sec. 8.2. For other choice at a different epoch (e.g. at an epoch in 2035 closer to eLISA/NGO planned arrival at science orbit), the procedure is the same.

**Table 2.** Initial states (conditions) of 3 S/C of eLISA/NGO at epoch JD2459215.5 (2021-Jan-1st 00:00:00) for our initial choice (third column), after 1st stage optimization (fourth column) and after all optimizations (fifth column) in J2000 equatorial (Earth mean equator and equinox coordinates) solar-system-barycentric coordinate system

| | | Initial choice of S/C initial states | Initial states of S/C after 1st stage optimization | Initial states of S/C after final optimization |
|---|---|---|---|---|
| S/C1 Position (AU) | X | $-1.53222193865\times10^{-2}$ | $-1.53222193865\times10^{-2}$ | $-1.53221933735\times10^{-2}$ |
| | Y | $9.23347976632\times10^{-1}$ | $9.23347976632\times10^{-1}$ | $9.23345222988\times10^{-1}$ |
| | Z | $4.04072005496\times10^{-1}$ | $4.04072005496\times10^{-1}$ | $4.04070800735\times10^{-1}$ |
| S/C1 Velocity (AU/day) | $V_x$ | $-1.71752389145\times10^{-2}$ | $-1.71926502995\times10^{-2}$ | $-1.71928071373\times10^{-2}$ |
| | $V_y$ | $-1.41699055355\times10^{-4}$ | $-1.41837311087\times10^{-4}$ | $-1.41838556464\times10^{-4}$ |
| | $V_z$ | $-6.11987395198\times10^{-5}$ | $-6.12586807155\times10^{-5}$ | $-6.12592206525\times10^{-5}$ |
| S/C2 Position (AU) | X | $-1.86344993528\times10^{-2}$ | $-1.86344993528\times10^{-2}$ | $-1.86344993528\times10^{-2}$ |
| | Y | $9.22658604804\times10^{-1}$ | $9.22658604804\times10^{-1}$ | $9.22658604804\times10^{-1}$ |
| | Z | $3.98334135807\times10^{-1}$ | $3.98334135807\times10^{-1}$ | $3.98334135807\times10^{-1}$ |
| S/C2 Velocity (AU/day) | $V_x$ | $-1.72244923440\times10^{-2}$ | $-1.72419907995\times10^{-2}$ | $-1.72419907995\times10^{-2}$ |
| | $V_y$ | $-1.88198725403\times10^{-4}$ | $-1.88384533079\times10^{-4}$ | $-1.88384533079\times10^{-4}$ |
| | $V_z$ | $-2.71845314386\times10^{-5}$ | $-2.72100311132\times10^{-5}$ | $-2.72100311132\times10^{-5}$ |
| S/C3 Position (AU) | X | $-1.19599845212\times10^{-2}$ | $-1.19599845212\times10^{-2}$ | $-1.19599845212\times10^{-2}$ |
| | Y | $9.22711604030\times10^{-1}$ | $9.22711604030\times10^{-1}$ | $9.22711604030\times10^{-1}$ |
| | Z | $3.98357113784\times10^{-1}$ | $3.98357113784\times10^{-1}$ | $3.98357113784\times10^{-1}$ |
| S/C3 Velocity (AU/day) | $V_x$ | $-1.72249891952\times10^{-2}$ | $-1.72424881557\times10^{-2}$ | $-1.72424881557\times10^{-2}$ |
| | $V_y$ | $-9.59855278460\times10^{-5}$ | $-9.60776184172\times10^{-5}$ | $-9.60776184172\times10^{-5}$ |
| | $V_z$ | $-9.55537821052\times10^{-5}$ | $-9.56487660660\times10^{-5}$ | $-9.56487660660\times10^{-5}$ |

### 7.2. Basic ASTROD orbit configuration

In the original proposal, the ASTROD-GW orbits are chosen in the ecliptic plane with inclination $\lambda = 0$. The angular resolution in the sky has antipodal ambiguity. Although over most of sky the resolution is good, near the ecliptic poles the resolution is poor. After 2010, we have designed the basic orbits of ASTROD-GW to have small inclinations in order to resolve these issues while keeping the variation of the arm lengths in the tolerable range.[39,40]

Following [39. 40], the basic idea is that if the orbits of the ASTROD-GW spacecraft are inclined with a small angle λ, the interferometry plane with appropriate design is also inclined with similar angle and when the ASTROD-GW formation evolves, the interferometry plane can be designed to modulate in the ecliptic solar-system barycentric frame. With this, angular positions of GW sources both near the polar region and off the polar region are resolved without antipodal ambiguity (see also Sec. 7.3).

Let first consider a circular orbit of a spacecraft in the Newtonian gravitational central problem (one-body central problem) in spherical coordinates ($r$, $\theta$, $\varphi$):

$$r = a,\ \theta = 90°,\ \varphi = \omega t + \varphi_0, \tag{26}$$

where $a$, $\omega$, and $\varphi_0$ are constants. For spacecraft in this discussion, we have $a$ = 1 AU, $\omega = 2\pi/T_0$ with $T_0$ = 1 sidereal year, and $\varphi_0$ is the initial phase in the coordinate considered. The spacecraft orbit at time $t$ in Cartesian coordinates is

$$x = a \cos \varphi = a \cos (\omega t + \varphi_0);\ y = a \sin \varphi = a \sin (\omega t + \varphi_0);\ z = 0. \tag{27}$$

Let transform this orbit actively into an orbit with inclination $\lambda$, and with the intersection of the orbit plane and $xy$-plane (the ecliptic) at the line $\varphi = \Phi_0$ in the $xy$-plane. The active transformation matrix is

$$\mathrm{R}(\lambda;\ \Phi_0) = \begin{pmatrix} \cos^2\Phi_0 + \sin^2\Phi_0\cos\lambda & \sin\Phi_0\cos\Phi_0(1 - \cos\lambda) & \sin\Phi_0\sin\lambda \\ \sin\Phi_0\cos\Phi_0(1 - \cos\lambda) & \sin^2\Phi_0 + \cos^2\Phi_0\cos\lambda & -\cos\Phi_0\sin\lambda \\ -\sin\Phi_0\sin\lambda & \cos\Phi_0\sin\lambda & \cos\lambda \end{pmatrix}. \tag{28}$$

The new spacecraft orbit is

$$\begin{pmatrix} x' \\ y' \\ z' \end{pmatrix} = \begin{pmatrix} a[1 - \sin^2\Phi_0(1 - \cos\lambda)]\cos\varphi + a \sin\Phi_0\cos\Phi_0(1 - \cos\lambda)\sin\varphi \\ a \cos\Phi_0\sin\Phi_0(1 - \cos\lambda)\cos\varphi + a[1 - \cos^2\Phi_0(1 - \cos\lambda)]\sin\varphi \\ -a \sin\Phi_0\sin\lambda\cos\varphi + a \cos\Phi_0\sin\lambda\sin\varphi \end{pmatrix}. \tag{29}$$

For the three orbits with inclination $\lambda$ (in radian), we choose:

S/C I: $\Phi_0(\mathrm{I}) = 270°$, $\varphi_0(\mathrm{I}) = 0°$;
S/C II: $\Phi_0(\mathrm{II}) = 150°$, $\varphi_0(\mathrm{II}) = 120°$;
S/C III: $\Phi_0(\mathrm{III}) = 30°$, $\varphi_0(\mathrm{III}) = 240°$. (30)

Defining

$$\xi \equiv 1 - \cos\lambda = 0.5\,\lambda^2 + O(\lambda^4), \tag{31}$$

from Eq. (29) and Eq. (30), we have

(i) for the orbit of S/C I

$$\begin{pmatrix} x^{\mathrm{I}} \\ y^{\mathrm{I}} \\ z^{\mathrm{I}} \end{pmatrix} = \begin{pmatrix} a\cos\omega t - \xi\, a\cos\omega t \\ a\sin\omega t \\ a\cos\omega t\sin\lambda \end{pmatrix}, \tag{32}$$

(ii) for the orbit of S/C II

$$\begin{pmatrix} x^{\mathrm{II}} \\ y^{\mathrm{II}} \\ z^{\mathrm{II}} \end{pmatrix} = \begin{pmatrix} a[(-\tfrac{1}{2})\cos\omega t - (3^{1/2}/2)\sin\omega t] + (a/2)\,\xi[(3^{1/2}/2)\sin\omega t - \tfrac{1}{2}\cos\omega t] \\ a[(-\tfrac{1}{2})\sin\omega t + (3^{1/2}/2)\cos\omega t] + (3^{1/2}/2)\,a\,\xi[(3^{1/2}/2)\sin\omega t - \tfrac{1}{2}\cos\omega t] \\ a\sin\lambda\,[(3^{1/2}/2)\sin\omega t - \tfrac{1}{2}\cos\omega t] \end{pmatrix}, \tag{33}$$

(iii) for the orbit of S/C III

$$\begin{pmatrix} x^{\mathrm{III}} \\ y^{\mathrm{III}} \\ z^{\mathrm{III}} \end{pmatrix} = \begin{pmatrix} a[(-\tfrac{1}{2})\cos\omega t + (3^{1/2}/2)\sin\omega t] + (a/2)\,\xi[(3^{1/2}/2)\sin\omega t - \tfrac{1}{2}\cos\omega t] \\ a[(-\tfrac{1}{2})\sin\omega t - (3^{1/2}/2)\cos\omega t] - (3^{1/2}/2)\,a\,\xi[(-3^{1/2}/2)\sin\omega t - \tfrac{1}{2}\cos\omega t] \\ a\sin\lambda\,[(-3^{1/2}/2)\sin\omega t - \tfrac{1}{2}\cos\omega t] \end{pmatrix}. \tag{34}$$

One can readily check that $[(x^{\mathrm{I}})^2 + (y^{\mathrm{I}})^2 + (z^{\mathrm{I}})^2]^{1/2} = [(x^{\mathrm{II}})^2 + (y^{\mathrm{II}})^2 + (z^{\mathrm{II}})^2]^{1/2} = [(x^{\mathrm{III}})^2 + (y^{\mathrm{III}})^2 + (z^{\mathrm{III}})^2]^{1/2} = a$ hold for consistency.

Calculate the arm vectors $\underline{V}_{\text{II-I}} = \underline{r}^{\mathrm{II}} - \underline{r}^{\mathrm{I}}$, $\underline{V}_{\text{III-II}} = \underline{r}^{\mathrm{III}} - \underline{r}^{\mathrm{II}}$ and $\underline{V}_{\text{I-III}} = \underline{r}^{\mathrm{III}} - \underline{r}^{\mathrm{I}}$:

$$\underline{V}_{\text{II-I}} = \begin{pmatrix} a[-(3/2)\cos\omega t - (3^{1/2}/2)\sin\omega t] + a\,\xi[(3^{1/2}/4)\sin\omega t + (3/4)\cos\omega t] \\ a[-(3/2)\sin\omega t + (3^{1/2}/2)\cos\omega t] + a\,\xi[(3/4)\sin\omega t - (3^{1/2}/4)\cos\omega t] \\ a\sin\lambda\,[(3^{1/2}/2)\sin\omega t - (3/2)\cos\omega t] \end{pmatrix}, \tag{35}$$

$$\underline{V}_{\text{III-II}} = \begin{pmatrix} 3^{1/2}\,a\sin\omega t - (3^{1/2}/2)\,a\,\xi\sin\omega t \\ -3^{1/2}\,a\cos\omega t + (3^{1/2}/2)\,a\,\xi\cos\omega t \\ -\,3^{1/2}\,a\sin\lambda\sin\omega t \end{pmatrix}, \tag{36}$$

$$\underline{V}_{\text{I-III}} = \begin{pmatrix} a[(3/2)\cos\omega t - (3^{1/2}/2)\sin\omega t] + a\,\xi[(3^{1/2}/4)\sin\omega t - (3/4)\cos\omega t] \\ a[(3/2)\sin\omega t + (3^{1/2}/2)\cos\omega t] + a\,\xi[-(3/4)\sin\omega t - (3^{1/2}/4)\cos\omega t] \\ a\sin\lambda\,[(3^{1/2}/2)\sin\omega t + (3/2)\cos\omega t] \end{pmatrix}, \tag{37}$$

The closure relation $\underline{V}_{\text{II-I}} + \underline{V}_{\text{III-II}} + \underline{V}_{\text{I-III}} = \underline{0}$ is checked for verifying calculations also. The arm lengths are calculated to be

$$|\underline{V}_{\text{II-I}}| = 3^{1/2}\,a\,[(1 - \xi/2)^2 + \sin^2\lambda\,\sin^2(\omega t - 60°)]^{1/2},$$

$$|\underline{V}_{\text{III-II}}| = 3^{1/2}\,a\,[(1 - \xi/2)^2 + \sin^2\lambda\,\sin^2(\omega t)]^{1/2},$$

$$|\underline{V}_{\text{I-III}}| = 3^{1/2}\, a\, [(1 - \xi/2)^2 + \sin^2\lambda \sin^2(\omega t + 60°)]^{1/2}. \tag{38}$$

The fractional arm length variation is within (1/2) $\sin^2\lambda$ which is about $10^{-4}$ for $\lambda$ around 1°.

The cross-product vector $\underline{N}(t) \equiv \underline{V}_{\text{III-II}} \times \underline{V}_{\text{I-III}}$ is normal to the orbit configuration plane and has the following components:

$$\underline{N} = [(3^{3/2}/2)\,(1 - \xi/2)\, a^2] \begin{pmatrix} -\sin\lambda \cos 2\omega t \\ -\sin\lambda \sin 2\omega t \\ (1 - \xi/2) \end{pmatrix}. \tag{39}$$

The normalized unit normal vector $\underline{n}$ is then:

$$\underline{n} = [\sin^2\lambda + (1 - \xi/2)^2]^{1/2} \begin{pmatrix} -\sin\lambda \cos 2\omega t \\ -\sin\lambda \sin 2\omega t \\ (1 - \xi/2) \end{pmatrix}. \tag{40}$$

The geometric center $\underline{V}_c$ of the ASTROD-GW spacecraft configuration is

$$\underline{V}_c = \begin{pmatrix} -(1/2)\,\xi\, a \cos\omega t \\ (1/2)\,\xi\, a \sin\omega t \\ 0 \end{pmatrix}. \tag{41}$$

There are 3 interferometers with 2 arms in the ASTROD-GW configuration. The geometric center of each of these 3 interferometers is at a distance of about 0.25 AU from the Sun. Numerical simulation and optimization of orbit configuration for inclination of 0.5°, 1°, 1.5°, 2°, 2.5° and 3° have been worked out using planetary ephemeris to take into account of the planetary perturbations in Ref. 41. The case with inclinations of 1° is reviewed in Sec. 8.3 for illustration. When LISA configuration orbits around the Sun, it is equivalent to multiple detector arrays distributed in 1 AU orbit. The extension of ASTROD-GW is already of 1 AU. When ASTROD-GW orbits around the Sun, it is also equivalent to multiple detector arrays distributed in 1 AU orbit.

### 7.3. Angular resolution

Consider angular resolution of a coherent GW source. Consider first the LISA case as example. The detector formation of LISA is modulated in its orbit around the Sun. The azimuth modulation amplitude is 2π rad with inclination 1.05 rad (60°) so that the antenna pattern sweeps around the sky in one year. The antenna response is not isotropic but the averaged linear angular resolution (in a year) of monochromatic GW sources for LISA differs by less than a factor of 3 among all directions.[9] This is also true for all LISA-like formations. *The angular resolution is basically proportional to the inverse of the strain signal to noise ratio.* If the inclination is of the order 0.017-0.052 rad (1-3°) for LISA, the polar resolution would be worsened by 30-10 times (approximately the ratio sin of 1.05 rad to sin of 0.03-0.1 rad); the steradian localization in the celestial sphere would be worsened by square of this factor. Away from the polar region ($\theta >> 0.017$-$0.052$ rad), the steradian localization in the celestial sphere would be by $\sin^2 \theta$. If the signal to noise ratio is downgraded by 5 (as in eLISA/NGO or in OMEGA in its low frequency part due to shorter arm length), the linear angular resolution is worsened by 5 times. ASTROD-GW has less sensitivity above 1 mHz compared with LISA, therefore the angular resolution will be worsened by both factors. In the 100 nHz-1 mHz region, ASTROD-GW has better sensitivity compared with LISA, in most part by 52 times. Hence, the angular resolution in the polar region is similar to that of LISA, while in other regions, the linear resolution is enhanced by roughly $52 \times \sin \theta$ (upgraded by 52 but downgraded by $\sin^2 \theta$ in sterad [by $\sin \theta$ in rad]). Although there is a mild dependence on the configuration inclination angle $\lambda$, within a factor of 3, the averaged antenna pattern for ASTROD-GW away from the polar region is better by a factor of $52 \times \sin \theta$ compared to that of LISA. Since the antenna pattern of ASTROD-GW sweeps over the whole sky in half year as can been seen from Eq. (40), the time of average needed is half a year instead of a year.[39]

For more complicated sources like chirping GW sources from BBHs (binary black holes), one needs to do fitting in order to obtain the accuracy of the parameters. However, the tendency of accuracy of parameters is the same: for similar situations, it is proportional to the inverse of the strain signal to noise ratio.

For Super-ASTROD, the strain signal to noise ratio would be even better than ASTROD-GW by 5 times toward the lower frequency region, therefore the angular resolution would be better by 5 times. For polar resolution, the ASTROD inclination strategy could be applied. However, since Super-ASTROD has 1 or 2 S/C in off-ecliptic orbit, this may not be needed. For ASTROD-EM, since the lunar orbit is inclined about 5° to the ecliptic and the node precession period is 18.61 tropical years, the Earth-Moon Lagrange points also precess together. Depending on the time and duration of mission, it might or might not be desirable to use slightly inclined orbit.[43]

Most of the Earth orbit GW missions have dipolar ambiguity and the resolution is poor in the polar region. However, this is not a big issue since we just needs to look at both polarity for identification of electromagnetic counterparts and the polar region is only a small portion of the sky.

### 7.4. Six/Twelve spacecraft formation

In order to detect relic GWs using correlated detection, Big Bang Observer[45] and DECIGO[44] proposals have 12 spacecraft distributed in the Earth orbit in 3 groups separated by 120 degrees in orbit; 2 groups has 3 spacecraft each in a LISA-like triangular formation and the third group has 6 spacecraft with two LISA-like triangles forming a star configuration (Fig. 7). An alternate configuration is that each group has 4 spacecraft forming a nearly square configuration (also has a tilt of 60 degrees with respect to the ecliptic plane).

For a more sensitive detection of background or relic GWs, correlated detection with 2 sets of triangular ASTROD-GW formation are required, i.e., a 6-S/C constellation. The second nearly triangular formation could be put again near L3, L4, L5 respectively, but separated from the first formation by $1 \times 10^6$ km to $5 \times 10^6$ km for the respective S/C.[40]

## 8. Orbit Design and Orbit Optimization using Ephemerides

Although Sun is dominant in the solar system, there are other planets and celestial objects affecting the orbit, notably Jupiter, Venus and Earth. Ephemerides is a must for orbit design. At present, there are 3 complete fundamental ephemerides of the solar system – DE (Development Ephemerides),[126] EPM (Ephemerides of Planets and Moon)[127] and INPOP (Intégrateur Numérique Planétaire de l'Observatoire de Paris).[128] Any of these 3 ephemerides could be used in the orbit design and orbit optimization. For easier in numerical processing, we normally use CGC (Center for Gravitation and Cosmology) ephemeris framework together with initial conditions taken from DE ephemerides at a certain epoch for evolving with post-Newtonian approximation.

### 8.1. CGC Ephemeris

In 1998, we started orbit simulation and parameter determination for ASTROD.[129,130] We worked out a post-Newtonian ephemeris of the solar system including the solar quadrupole moment, the eight planets, the Pluto, the moon and the 3 biggest asteroids. We term this working ephemeris CGC 1 (CGC: Center for Gravitation and Cosmology). Using this ephemeris as a deterministic model and adding stochastic terms to simulate noise, we generate simulated ranging data and use Kalman filtering to determine the accuracies of fitted relativistic and solar-system parameters for 1050 days of the ASTROD mission.

For a better evaluation of the accuracy of $\dot{G}/G$, we need also to monitor the masses of other asteroids. For this, we considered all known 492 asteroids with diameter greater than 65 km to obtain an improved ephemeris framework --- CGC 2, and calculated the perturbations due to these 492 asteroids on the ASTROD spacecraft.[131,132]

In building CGC ephemeris framework, we use the post-Newtonian barycentric metric and equations of motion as derived in Brumberg[133] with PPN (Parametrized Post-Newtonian) parameters $\beta$ and $\gamma$ for solar system bodies (with the gauge parameter $\alpha$ set to zero). These equations are used to build our computer-integrated ephemeris (with the PPN parameters $\gamma = \beta = 1$, $J_2 = 2 \times 10^{-7}$) for eight-planets, the Pluto, the Moon and the Sun. The initial positions and initial velocities at the epoch 2005.6.10 0:00 are taken from the DE403 ephemeris. The evolution is solved by using the 4$^{th}$-order Runge-Kutta method with the step size h =0.01 day. In Ref. 130, the 11-body evolution is extended to 14-body to include the 3 big asteroids — Ceres, Pallas and Vesta (CGC 1 ephemeris). Since the tilt of the axis of the solar quadrupole moment to the perpendicular of the elliptical plane is small (7°), in CGC 1 ephemeris, we have neglected this tilt. In CGC 2 ephemeris, we have added the perturbations of additional 489 asteroids.

In our first optimization of ASTROD-GW orbits,[134-136] we have used CGC 2.5 ephemeris in which only 3 biggest minor planets are taken into accounts, but the Earth's precession and nutation are added; the solar quadratic zonal harmonic and the Earth's quadratic to quartic zonal harmonic are considered. In later simulation, we add the perturbation of additional 349 asteroids and call it CGC 2.7 ephemeris.[85-87] The differences in orbit evolution compared with DE405 for Earth for 3700 days starting at JD2461944.0 (2028-Jun-21 12:00:00) are shown in Fig. 5 of Ref. 40. The differences in radial distances are less than about 200 m. The differences for other inner planets are smaller. The differences in latitude and longitude for Earth are less than 1 mas.

### 8.2. Numerical orbit design and orbit optimization for eLISA/NGO

The mission orbit configuration of eLISA/NGO is similar to that of LISA but with a shorter arm length and a closer distance to Earth. The distance of any two of three spacecraft must be maintained as close as possible during geodetic flight. LISA orbit configuration has been studied analytically and numerically in various previous works.[118-123,137,138] For eLISA/NGO, we followed the analytical procedure of Dhurandhar *et al.*[123] [see Sec. 7.1] in making our initial choice of the initial conditions in Ref. [85]; these initial conditions are listed in column 3 of Table 2 in Sec 7.1. With this orbit choice, we started numerical orbit design and used the CGC ephemeris to numerically optimize the orbit configuration in [85] as we have done for ASTROD-GW orbit[134-136,139,86,87] design. In this subsection, we review and summarize the procedure following [85].

The goal of the eLISA/NGO mission orbit optimization is to equalize the three arm lengths of the eLISA/NGO formation and to reduce the relative line-of-sight velocities between three pairs of spacecraft as much as possible. In the solar system, the eLISA/NGO spacecraft orbits are perturbed by the planets. With the initial states of the three spacecraft as listed in column three of table 2, we calculated the eLISA/NGO orbit configuration for 1000 days using CGC 2.7. The variations of arm lengths and velocities in the line of sight direction are drawn in figure 2 of [85]. The largest variations are caused by Earth, Jupiter and Venus. Our method of optimization is to modify the initial velocities and initial heliocentric distances so that (i) the perturbed orbital periods for 1000-day average remains close to one another, and (ii) the average major axes are adjusted to make arms nearly equal. We do this iteratively as follows. From figure 2 of [85], we noticed that the variation of Arm1 (between S/C2 and S/C3) is small. First, we adjust the initial conditions of S/C2 and S/C3 to make the variation of Arm1 satisfy the mission requirements that arm length variations are within 2 % and Doppler velocities are within 10 m/s. Then we adjust the initial conditions of S/C1 so that Arm2 and Arm3 satisfy the mission requirements. Adjustments are always performed in the ecliptic heliocentric coordinate system.

The actual adjustment procedure is described as follows. Firstly, the magnitudes of initial velocities of S/C2 and S/C3 were adjusted so that their average periods (367.474 days) in 3 years were a little bit longer than 1 sidereal year. Within a definite range, when the periods become longer, the variations of Arm1 become smaller. The initial velocities were adjusted so that the Arm1 satisfied the eLISA/NGO arm length and Doppler velocity requirements. After this, we adjusted the initial velocities of S/C1 to make its orbital period approach those of S/C2 and S/C3, and Arm2 and Arm3 nearly equal. If the results obtained from the above procedure did not satisfy the requirements or better results were expected, we could adjust the orbital periods of S/C2 and S/C3 a little bit longer again under the constraint that the eLISA/NGO requirements for Arm1 is satisfied. Up to this stage, only initial velocities have been adjusted. After we have completed this stage, the initial conditions of the 3 S/C are listed in column 4 of Table 2; the variations of arm lengths and velocities in the line of sight direction are drawn in figure 3 of [85].

After the first stage, we optimized the orbital period of S/C1 by adjusting the initial velocity and the semi-major axis until the eLISA/NGO requirements were satisfied. The initial conditions of the 3 S/C, after optimization, are listed in column 5 of Table 2; the variations of arm lengths (within 2 %) and velocities in the line of sight direction (within 5.5 m/s) are drawn in figure 8. In figure 8, we also draw the angle between S/C and Earth subtended from Sun in 1000 days; it starts at 10° behind Earth (in solar orbit) and varies between 9° and 16° with a quasi-period of variation about 1 sidereal year mainly due to Earth’s elliptic motion.

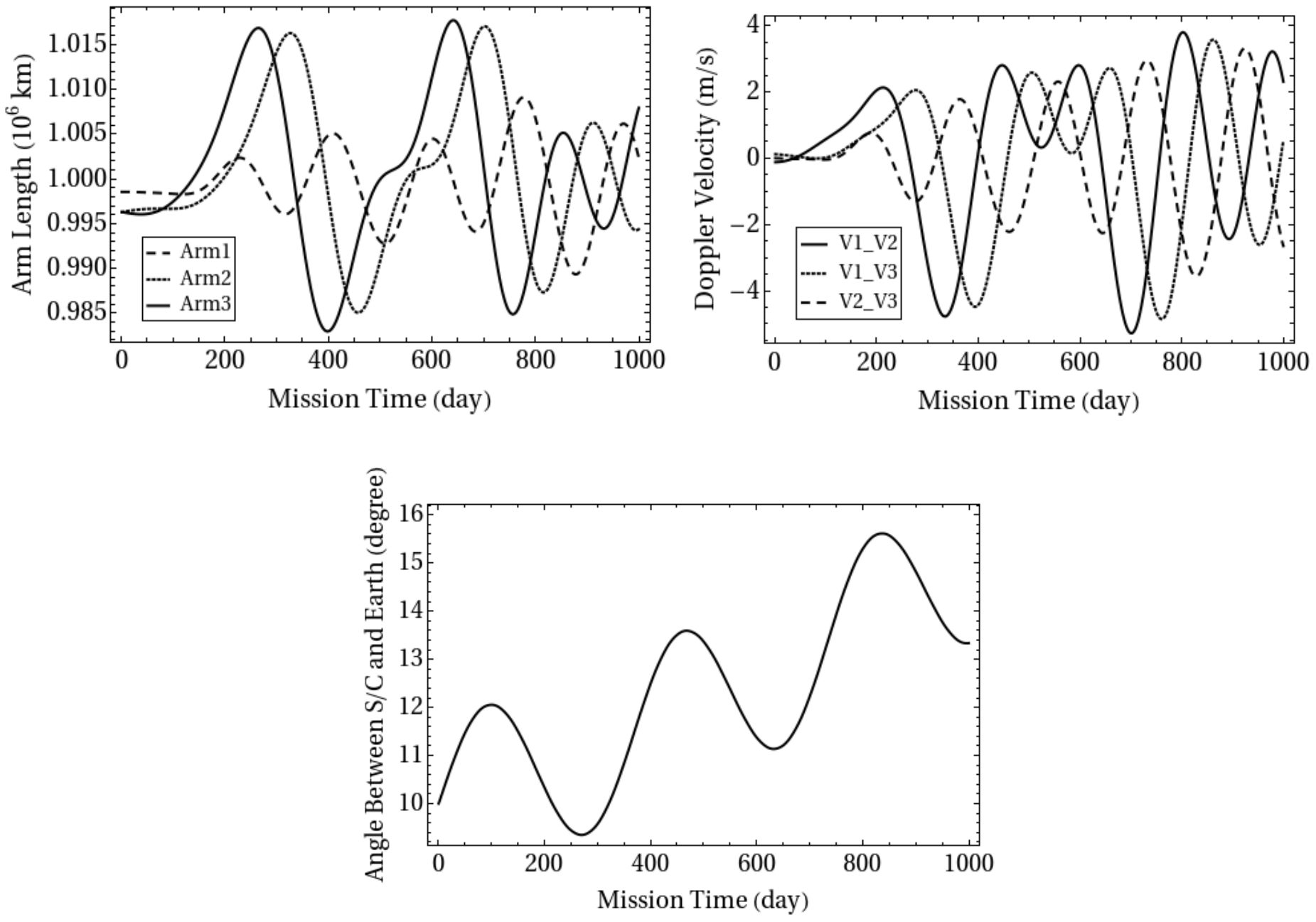

**Figure 8.** Variations of the arm lengths, the velocities in the line of sight direction, and the angle between S/C and Earth subtended from Sun in 1000 days for the S/C configuration with initial conditions given in column 5 (after final optimization) of Table 2.

### 8.3. Orbit Optimization for ASTROD-GW[41]

The goal of the ASTROD-GW mission orbit optimization is to equalize the three arm lengths of the ASTROD-GW formation and to reduce the relative line-of-sight velocities between three pairs of spacecraft as much as possible. In our first optimization, the time of start of the science part of the mission is chosen to be noon, June 21, 2025 (JD2460848.0) and the optimization is for a period of 3700 days using CGC 2.5 ephemeris.[134-136] Since the preparation of the mission may take longer time and there is a potential that the extended mission life time may be longer than 10 years, in later optimizations,[139,86,87] we started at noon, June 21, 2028 (JD2461944.0) and optimize for a period of 20 years using CGC 2.7 ephemeris including more asteroids than those of CGC 2.5. In both of these optimizations, the orbit configuration is set in the ecliptic plane and we have the inclination angle $\lambda = 0$. With the basic configuration of ASTROD-GW changed into an inclined precession orbit formation, we re-design and re-optimize our orbit configuration numerically starting at noon, June 21, 2035 (JD2464500.0) for 10 years for the inclination angle 0.5°, 1°, 1.5°. 2°, 2.5°and 3° using the CGC 2.7.1 ephemeris.[41]

In this subsection, we illustrate the design and optimization method with inclined precession orbit formation for the case having inclination angle 1° following [41] which uses CGC 2.7.1. The differences between CGC 2.7.1 and CGC 2.7 (summarized in Sec. 8.1) is detailed in subsection 8.3.1. In subsection 8.3.2, we review how to obtain the initial choice of S/C initial conditions as a starting point for numerical optimization. In subsection 8.3.3, we discuss method of optimization and summarize the results of optimization.

### 8.3.1 CGC 2.7.1 Ephemeris

In the CGC 2.7.1 ephemeris framework, we pick up 340 asteroids besides the Ceres, Pallas and Vesta from the Lowell database. The masses of 340 asteroids are given by Lowell data[140] instead of estimating the masses based on the classification in CGC 2.7.[84,85,87] The orbit elements of these asteroids are also updated from the Lowell database.

For a 10-year duration starting at June 21, 2035, the differences between the Earth's heliocentric distances calculated by CGC 2.7.1 and DE430 are within 150 m, and that the differences in longitudes and latitudes are within 1.4 mas and 0.45 mas respectively. These differences do not affect the results of our TDI calculations.

### 8.3.2 Initial choice of spacecraft initial conditions

The R.A. of the Earth at JD2464500 (2035-June-21st 12:00:00) is $17^h57^m45.09^s$, i.e. 269.438° from DE 430 ephemeris. The initial positions of the 3 S/Cs are obtained by choosing the $\omega t$ as 89.44° for $\varphi = \omega t + \varphi_0$ in Equation (29). The initial velocities are derived from Equation (29) by calculating the derivatives with respect to $t$. The S/C1 orbit near the Lagrange point L3 is partly obscured by Sun from the line of sight of Earth (left diagram of Fig. 9). It would obstruct the communication with the Earth stations. To avoid the obscuration, we rotate the initial angle $\Phi_0$ and $\varphi_0$ forward of by 2.0° for inclination angle 1.0°. The S/C1 orbit is shown on the right diagram of Fig. 9. The initial choice of initial states for the 3 S/Cs in this case is listed in column 3 of Table 3.

### 8.3.3 Method of optimization

Our optimization method is to modify the initial velocities and initial heliocentric distances to reach the aim of (i) equalizing the three arm lengths of the ASTROD-GW formation as much as possible and (ii) reducing the relative Doppler velocities between three pairs of spacecraft as much as possible.

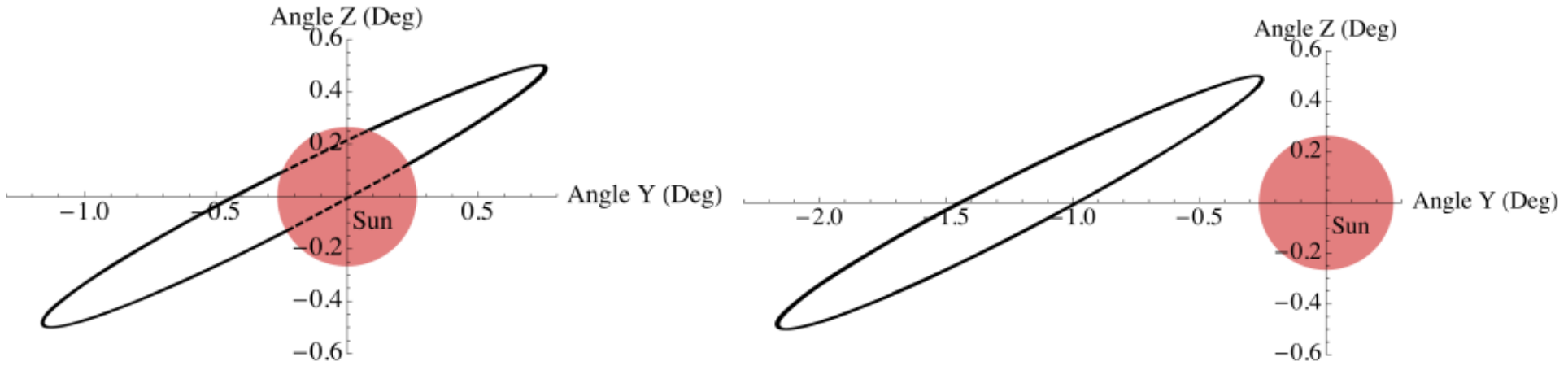

**Fig. 9.** S/C1 view from Earth before rotating the initial conditions by an angle (left diagram) and after rotating by an angle 2.0° (right diagram) for the case of inclination angle 1.0°.[41]

**Table 3.** Initial states of S/Cs for the configuration with the inclination angle 1° at epoch JD2464500.0 for initial choice, after period optimization, and after all optimizations in J2000 equatorial solar-system-barycentric coordinate system.[41]

| $\lambda$ = 1.0° | | Initial choice of S/C initial states | Initial states of S/Cs after period optimization | Initial states of S/C after final optimization |
|---|---|---|---|---|
| S/C1 | X | -2.8842263289715×10$^{-2}$ | -2.8842263289715×10$^{-2}$ | -2.8842514605546×10$^{-2}$ |
| Position | Y | 9.1157742309044×10$^{-1}$ | 9.1157742309044×10$^{-1}$ | 9.1158659433458×10$^{-1}$ |
| (AU) | Z | 3.9552690922456×10$^{-1}$ | 3.9552690922456×10$^{-1}$ | 3.9553088730467×10$^{-1}$ |
| S/C1 | $V_x$ | -1.7188548244458×10$^{-2}$ | -1.7188535691176×10$^{-2}$ | -1.7188363750567×10$^{-2}$ |
| Velocity | $V_y$ | -2.8220395391983×10$^{-4}$ | -2.8220375159556×10$^{-4}$ | -2.8220098038726×10$^{-4}$ |
| (AU/day) | $V_z$ | -4.4970276654173×10$^{-4}$ | -4.4970243993363×10$^{-4}$ | -4.4969796642665×10$^{-4}$ |
| S/C2 | X | 8.7453598387569×10$^{-1}$ | 8.7453598387569×10$^{-1}$ | 8.7453598387569×10$^{-1}$ |
| Position | Y | -4.3802677355114×10$^{-1}$ | -4.3802677355114×10$^{-1}$ | -4.3802677355114×10$^{-1}$ |
| (AU) | Z | -2.0634980179207×10$^{-1}$ | -2.0634980179207×10$^{-1}$ | -2.0634980179207×10$^{-1}$ |
| S/C2 | $V_x$ | 8.2301784322477×10$^{-3}$ | 8.2301033726700×10$^{-3}$ | 8.2301033726700×10$^{-3}$ |
| Velocity | $V_y$ | 1.3797379424198×10$^{-2}$ | 1.3797253460590×10$^{-2}$ | 1.3797253460590×10$^{-2}$ |
| (AU/day) | $V_z$ | 6.1425805519808×10$^{-3}$ | 6.1425244722884×10$^{-3}$ | 6.1425244722884×10$^{-3}$ |
| S/C3 | X | -8.5683596527799×10$^{-1}$ | -8.5683596527799×10$^{-1}$ | -8.5679330969623×10$^{-1}$ |
| Position | Y | -4.8998222347472×10$^{-1}$ | -4.8998222347472×10$^{-1}$ | -4.8995800210059×10$^{-1}$ |
| (AU) | Z | -1.9592963105165×10$^{-1}$ | -1.9592963105165×10$^{-1}$ | -1.9591994878015×10$^{-1}$ |
| S/C3 | $V_x$ | 8.9788714330506×10$^{-3}$ | 8.9787977300014×10$^{-3}$ | 8.9792464008067×10$^{-3}$ |
| Velocity | $V_y$ | -1.3530263187520×10$^{-2}$ | -1.3530152097744×10$^{-2}$ | -1.3530828362023×10$^{-2}$ |
| (AU/day) | $V_z$ | -5.6998631854817×10$^{-3}$ | -5.6998163886731×10$^{-3}$ | -5.7001012664635×10$^{-3}$ |

During the actual optimization procedure, we use the following equation to modify the average period of the orbit:

$$V_{new} = V_{prev} + \Delta V \approx \left(1 - \frac{1}{3}\frac{\Delta T}{T}\right) V_{prev}. \tag{42}$$

For the case of inclination angle of 1°, we calculate the 3 S/Cs orbits with the initial choice of initial conditions listed in column 3 of Table 3 using the CGC 2.7.1 ephemeris. The average periods of the 3 S/Cs in 10 years are 365.256 days (S/C1), 365.267 days (S/C2) and 365.266 days (S/C3) respectively. We use equation (42) to change the initial velocities so that the average period of S/C1, S/C2 and S/C3 is adjusted to 365.255 days, 365.257 days and 365.257 days respectively. The initial conditions after this step are listed in column 4 of Table 3. In the next step, we use the following equations to trim the S/C eccentricities to be nearly circular:

$$R_{\rm new} = R_{\rm prev} + \Delta R \approx \left(1 + \frac{\Delta R}{R}\right) R_{\rm prev},$$

$$V_{\rm new} = V_{\rm prev} + \Delta V \approx \left(1 - \frac{\Delta R}{R}\right) V_{\rm prev}. \quad (43)$$

Here $R$ is the initial heliocentric distance of spacecraft. The fractional adjustment $\pm(\Delta R/R)$ in $\boldsymbol{R}_{prev}$ and $\boldsymbol{V}_{prev}$ would adjust eccentricity without adjust the period of the orbit. The initial conditions after all optimization are listed in column 5 of Table 3.

For the inclination angles 0.0°, 0.5°, 1.5°, 2°, 2.5° and 3°, the optimization processes are similar to the inclination 1.0° and the results can be found in [41].

## 9. Deployment of Formation in Earth-like Solar Orbit

The deployment to orbit around Earth, to halo orbit of Earth-Moon Lagrange points and to Sun-Earth L1 and L2 points are well studied. Here we say a few words on the deployment of a spacecraft to different positions of an Earth-like solar orbit. A preliminary design of the transfer orbits of the spacecraft from the separations of the launch vehicles to the mission orbits near L3, L4 and L5 points has been given in Ref. [40, 141]. Let us review this preliminary design first.

In the mission study of ASTROD I, the ASTROD I S/C is given an appropriate delta-V before the last stage of launcher separation in the LEO (Low Earth Orbit) and is injected directly to the solar orbit going geodetic to Venus swing-by. We can use the same strategy to launch the ASTROD-GW S/C directly into the solar transfer orbits near the designated Hohmann transfer orbits or Venus swing-by orbit. This way, the only major delta V needed for each S/C to reach the destination occurs near the destination to boost the S/C to stay near the destined Lagrange point. In row 2-4 of Table 4, we list types of transfer orbits, transfer times, the values of solar transfer delta-V and propellant mass ratio for 3 ASTROD-GW S/C. These estimates are good for any other S/C deployed to the same positions. The propellant mass ratios are around 0.5-0.55, 0.280 and 0.47 for S/C 1, 2 and 3. The total masses in case of ASTROD-GW S/C correspond to a dry mass of 500 kg are 1111-1266 kg, 723 kg, and 1035 kg for 3 S/C respectively (including the propellant and the propulsion module with mass of 10% of the propellant).

Table 4. Estimated Delta-V and Propellant Mass Ratio for Solar transfer of S/C

| ° ahead of Earth in solar orbit | Transfer Orbit | Transfer Time | Solar Transfer Delta-V after injection from LEO to solar transfer orbit | Solar Transfer Propellant Mass Ratio (Isp=320 s) |
|---|---|---|---|---|
| 180° (near L3) | Venus flyby transfer | 1.3-1.5 yr | 2.2-2.5 km/s | 0.50-0.55 |
| 60° (near L4) | Inner Hohmann, 2 Revolutions | 1.833 yr | 1.028 km/s | 0.280 |
| 300° (− 60°) (near L5) | Outer Hohmann, 1 Revolutions | 1.167 yr | 2 km/s | 0.47 |
| 0 - 60° | Inner Hohmann, ≤ 2 Revolutions | Less than 1.833 yr | Less than 1.028 km/s | Larger than 0.280 |
| 60° - 300° | Venus flyby transfer | 1.3-1.5 yr | 2.2-2.5 km/s | 0.50-0.55 |
| 300° - 360° | Outer Hohmann, 1 Revolutions | 1.167 yr | 2 km/s | 0.47 |

For deployment to other location in the solar orbit, we made estimates and list them in row 5-7 of Table 4. The baseline is: (i) S/C is propelled by a high efficient propulsion module (including the propellant with specific impulse 320 sec and the propulsion module with mass of 10% of the propellant) for large delta-V maneuvers and for delivery to the destination; (ii) This module is to be separated when the destination state is achieved.

Further studies on the optimizations of deployment from separation of launcher(s) for the orbit configurations with inclinations and for a period of 20 years are ongoing for both LISA-like missions and ASTROD-GW like missions.[142]

## 10. Time Delay Interferometry (TDI)

In Section 4 we start discussing TDI, now we continue. *To achieve required GW sensitivity, TDI to suppress laser frequency noise is required for space GW missions.*

Schematic orbit configuration of LISA-type mission design[9] and ASTROD-GW mission design[41] are shown in Fig. 1 and Fig. 2 respectively. For the numerical evaluation, we take a common receiving time epoch for both beams; the results would be very close to each other numerically if we take the same start time epoch and calculate the path differences. We refer to the path S/C1→S/C2→S/C1 as *a* (path) and the path S/C1 → S/C3 → S/C1 as *b* (path). Hence the difference $\Delta L$ between Path 1 and Path 2 for the unequal-arm Michelson can be denoted as $ab - ba \equiv [a,b]$. Here *ab* means *a* path followed by *b* path. The unequal-arm Michelson is now commonly called X-configuration.[88,89] The result of this TDI calculation for ASTROD-GW orbit with 1º inclination is shown in Fig. 10.

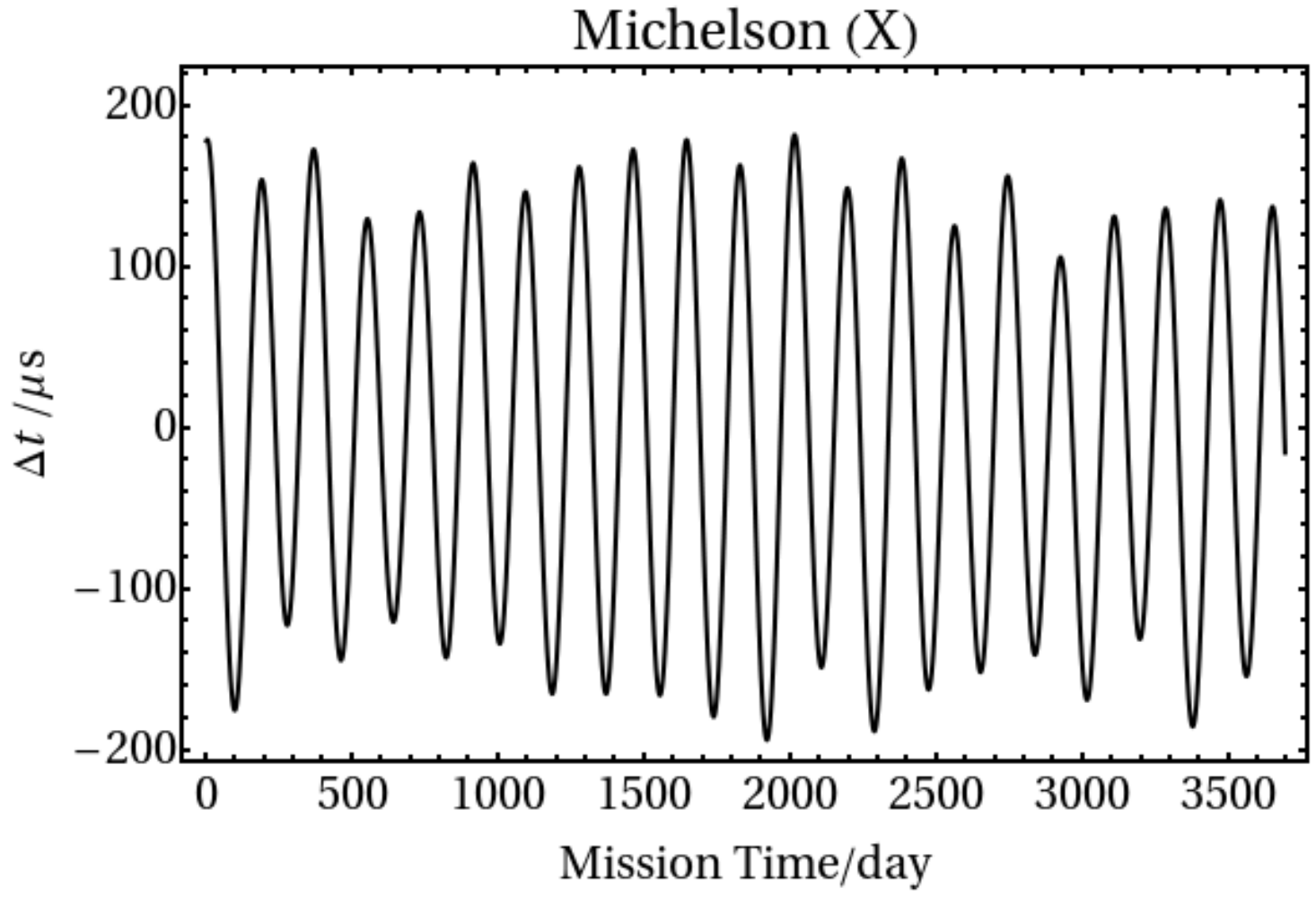

**Fig. 10.** Path length differences between two optical paths of the Unequal-arm Michelson TDI configuration (X-configuration) for ASTROD-GW orbit formation with 1º inclination.

The first-generation and second-generation TDIs are proposed since 1999.[88,89] In the first generation TDIs, static situations are considered. While in the second generation

TDIs, motions are compensated to a certain degree. The X configurations considered above belong to the first generation TDI configurations. We note that the numerical method has the advantage of taking care of all generations into a single calculation format. We shall not review more about these developments here, but refer the readers to the excellent review article by Tinto and Dhurandhar[89] for a comprehensive treatment.

We compile for comparison the resulting differences for second generation two-arm TDIs with n=1 and n=2 (n is the degree of polynominal in *ab*, *ba*, $a^2$ and $b^2$) due to arm length variations for various mission proposals -- eLISA/NGO, an NGO-LISA-type mission with a nominal arm length of 2 × $10^6$ km, LISA and ASTROD-GW in Table 5.

**Table 5.** Comparison the resulting differences for second generation TDIs (n=1 and n=2) due to arm length variations for various mission proposals -- eLISA/NGO, an NGO-LISA-type mission with a nominal arm length of 2 × $10^6$ km, LISA and ASTROD-GW.

| TDI configuration | | TDI path difference $\Delta L$ | | | |
|---|---|---|---|---|---|
| | | eLISA/NGO [85] | NGO-LISA-type with 2 × $10^6$ km arm length [85] | LISA [84] | ASTROD-GW (1° inclination) [41] |
| Duration | | 1000 days | 1000 days | 1000 days | 10 years |
| n=1 | [*ab*, *ba*] | -1.5 to +1.5 ps | -11 to +12 ps | -70 to +80 ps | -228 to +228 μs |
| n=2 | [$a^2b^2$, $b^2a^2$] | -11 to +12 ps | -90 to +100 ps | -600 to +650 ps | -1813 to +1813 ns |
| | [*abab*, *baba*] | -6 to +6 ps | -45 to +50 ps | -300 to +340 ps | -907 to +907 ns |
| | [$ab^2a$, $ba^2b$] | -0.0032 to +0.0034 ps | -0.0036 to +0.004 ps | -0.015 to +0.013 ps | -0.66 to +0.66 ns |
| Nominal arm length | | 1 Gm (1 Mkm) | 2 Gm | 5 Gm | 260 Gm |
| Requirement on $\Delta L$ | | 10 m (30 ns) | 20 m (60 ns) | 50 m (150 ns) | 500 m (1,500 ns) |

We note that:

(i) All the second-generation TDIs considered for the one-detector case for eLISA/NGO, for NGO-LISA-type with 2 × $10^6$ km arm length, for LISA and for ASTROD-GW with 1° inclination basically satisfy the requirement. (Table 5)

(ii) The requirement for unequal arm Michelson (X-configuration) TDI of ASTROD-GW needs to be relaxed by about 2 orders. (Fig. 10 and Table 5).

(iii) In view of the possibility of a GW mission in Earth orbit, numerical TDI study for GW missions in Earth orbit are desired.

(iv) Experimental demonstration of TDI in laboratory for LISA has been implemented in 2010.[143] eLISA and the original ASTROD-GW TDI requirement are based on LISA requirement, and hence also demonstrated. With the present pace of development in laser technology, the laser frequency noise requirement is expected to be able to compensate for 2-3 order of TDI requirement relaxation in 20 years.

(v) X-configuration TDI sensitivity for GW sources has been studied extensively for eLISA.[21] It satisfies the present technological requirements well. With enhanced laser technology expected, it would also be good for studying the ASTROD-GW and various GW missions in Earth orbit. The study for GW sensitivity and GW sources for other first-generation and second-generation TDIs and for other missions would also be encouraged.

## 11. Payload Concept

GW detection in space basically measures the distance change between two S/C (or celestial bodies) as GW comes by. The two S/C (or celestial bodies) must be in geodesic motion (or such motion can be deduced). The distance measurement must be ultra-sensitive as the GWs are weak. A typical implementation (mission) consists of three spacecraft in an almost equilateral triangle formation. The 3 spacecraft range interferometrically with one another. Each spacecraft carries a payload of two proof masses, two telescopes, two lasers, a weak light detection and handling system, a laser stabilization system, and a drag-free system. For lower part of space GW band or for possibly higher precision, a precision/optical clock, or an absolute laser stabilization system, and an absolute laser metrology system may be used.

*Weak light phase locking and handling*: For solar orbit missions, this is important. For ASTROD-GW with a distance of 260 Gm (1.73 AU), there is a need to phase lock a local laser to 100 fW incoming light to amplify and manipulate it. For 100 fW ($\lambda$ = 1064 nm) weak light, there are $5 \times 10^5$ photons/s. This would be good for 100 kHz frequency tuning. For LISA, 85 pW weak light phase locking is required. In Tsing Hua University, 2 pW weak-light phase-locking with 0.2 mW local oscillator has been demonstrated.[29,30] In JPL (Jet Propulsion Laboratory), Dick *et al*.[83] have achieved offset phase locking to 40 fW incoming laser light. It would be good for future development focusing on frequency-tracking, modulation-demodulation and coding-decoding to make it a mature experimental technique. This is also important for the deep space optical communication.

*Drag-free system design and development*: Drag-free system consists of a high precision accelerometer/inertial sensor to detect non-drag-free motions and a micro-thruster system to do the feedback to keep the spacecraft drag-free. LISA Pathfinder successfully demonstrated and tested the drag-free technology in the frequency range above 100 μHz to satisfy not just the requirement of LISA Pathfinder, but also the requirement of LISA.[11] The success paved the road of knowledge for all the space mission proposal in Table I. However, for lower part (100 nHz to 100 μHz) of the space frequency band, there needs more work. We have discussed frequency sensitivity spectrum and reddening factors in Sec, 5. To suppress the reddening factors requires position sensing noise to be flat down to 100 nHz and gravity acceleration due to spacecraft to be small and modeled to the required level at low frequencies. The self-gravity-acceleration needs to be stable or subtracted in real time. An absolute laser metrology system to monitor positions of major mass distribution in the S/C will be implemented to do this. To completely drop the factor or to go beyond, one may need to go to optical sensing and optical feedback control. As to the accelerometer/inertial sensor design of ASTROD, an absolute laser metrology system is proposed to push the noise down, in particular in the lower frequency region. In addition, ASTROD is proposed to monitor the positions of various parts of the spacecraft, to facilitate gravitational modeling.[27,28]

*Micro-thruster system*: For drag-free feedback control, micro-thrusters are needed. Field Emission Electric Propulsion (FEEP) system with its high specific thrust is a good candidate for the micro-thruster system. The sensitivity of FEEP system is good and is in the μN range. The main issue for FEEPs is lifetime. Due to technical problems during the development of the FEEP technology, the cold gas thrusters have become the alternative choice. The GAIA mission carries cold gas thrusters for the AOCS (Attitude and orbit control system).[144] MICROSCOPE[145] and LISA Pathfinder are equipped with cold gas thrusters based on the GAIA thrusters. The main disadvantage of cold gas thrusters compared to FEEPs is the higher mass per delta-V. The total mission duration is limited by the amount of propellant stored in the tanks. Therefore the FEEP technology would be preferred if it is available at a later time.

*Laser system*: Nd:YAG Non-planar ring oscillators pumped by laser diodes are available with output power of 2 W for use. The frequency noises must suppressed to very low level. The strategy is like the one adopted by NGO/eLISA using pre-stabilization, arm locking[21] and TDI (Sec. 10).

*Laser frequency standard/Clock*: Space optical clocks and optical comb frequency synthesizer technologies are important in the realization and simplification of the GW mission target sensitivity at lower frequency. Another use of the optical clock and optical comb frequency synthesizer is to calibrate the optical metrology for ASTROD-GW like missions. This is important for the laser metrology inertial sensor and for monitoring distances inside spacecraft, to correct local gravity changes due to, for example, thermal effects. All these measurements use lasers as standard rods. They need to be calibrated using optical frequency standards or absolutely stabilized laser frequency standard referenced to an atomic or molecular line. The advent of optical clocks and optical combs in space may possibly simplify the experimental design of ASTROD-GW like mission.

At present, optical clocks in the laboratory[146] have reached a fractional inaccuracy at $10^{-18}$ level; and they are improving. Clocks of this accuracy level or better can be used for exquisitely sensitive measurements of gravity, motion, and inertial navigation. The use of this kind of clocks certainly will facilitate the detection of the lower frequency GWs and stimulate the needs of re-design the implementation schemes of the lower frequency space GW detection.

*Absolute laser metrology system*: With an ultraprecise laser frequency standard/clock, an absolute laser metrology system can be built to monitor the positions of various parts of the spacecraft to facilitate gravitational modeling.

*Radiation monitor*: A small radiation detector onboard the spacecraft will monitor test-mass charging of the inertial drag-free sensors. This radiation monitor can also be used for measuring solar energetic particles (SEPs) and galactic cosmic rays (GCRs) in the area of solar and galactic physics with corresponding applications to space weather.[147,148]

## 12. Outlook

White dwarf was discovered in 1910 with its density soon estimated. Now we understand that GWs from white dwarf binaries in our Galaxy form a stochastic GW background (“confusion limit”)[149] for space GW detection in general relativity. The characteristic strain for confusion limit is about $10^{-20}$ in 0.1-1 mHz band. As to individual sources, some can have characteristic strain around this level for frequency 1-3 mHz in low-frequency band. One hundred year ago, the sensitivity of astrometric observation through the atmosphere around this band is about 1 arcsec. This means the strain sensitivity to GW detection is about $10^{-5}$; 15 orders away from the required sensitivity.

The first artificial satellite Sputnik was launched in 1957. The technological demonstration mission LISA Pathfinder was launched on 3 December, 2015. This mission successfully tested and demonstrated the drag-free technology to satisfy not just the requirement of LISA Pathfinder, but also basically the drag-free requirement of LISA GW space mission concept.[11] Thus, the major issue in the technological gap of 15 orders of magnitude is successfully abridged during last hundred years. The success paved the road for all the space mission proposals (Table 1). At present the space GW missions are expected to be launched in two decades. Weak-light phase locking is demonstrated in laboratories.[29,30,83] Weak-light technology still needs developments. And we do anticipate the possibility of an earlier launch date for eLISA (or a substitute mission) and possible earlier flight of other missions. With the first direct detection of GWs by LIGO and the success of LISA Pathfinder mission, the outlook of space detection of GWs is bright.

The science goals of space GW detectors are the detection of GWs from (i) Massive Black Holes; (ii) Extreme-Mass-Ratio Black Hole Inspirals; (iii) Intermediate-Mass Black Holes; (iv) Galactic Compact Binaries and (v) Relic GW Background. As we can readily see from Fig.'s 4-6, the signal-to-noise ratios (S/N) for GW detection of MBHB mergers are very high, and for the high S/N detection of more massive mergers the strain sensitivity at lower part (100 nHz – 100 μHz) of the space detection band is important. For doing this, longer arms have advantages. Longer arm missions would be good to compliment PTAs in the exploration of black hole co-evolution with galaxies. Longer arm missions with its better angle resolution are also more effective in the determination of the equation of state of dark energy, testing relativistic gravity and, possibly, probing the inflationary physics. Efforts in minimizing the accelerometer/inertial sensor noise over the MLDC formula or beyond will strengthen these goals. Deployment of S/C to any position in the Earth-like solar orbit could be less than 1.8 years with propellant mass ratio less than 0.55. This is within the practical range of launcher implementation.

Now we list important issues for further studies in order to realize and sharpen our expectations for GW detection in the frequency range 100 nHz – 100 μHz:

(i) Manipulating weak light;
(ii) Improvement of low-frequency acceleration noise;
(iii) Fourier spectrum of perturbations due to celestial bodies in the solar system and the precision needed to know the positions of solar-system bodies in order to separate this spectrum from GW spectrum;
(iv) Further studies in optimizing deployment delta-V and propellant ratio;
(v) Optimizing the inclination angle of the ASTROD-GW like constellation;
(vi) Extraction of GW signals based on precise numerical orbits;
(vii) Further studies in the angular resolution of GW sources;
(viii) Separation of weak lensing effects from GW signals.

It is time to think seriously about second-generation space GW detectors – BBO, DECIGO, Super-ASTROD and the like. Optical clocks in the laboratory have reached a fractional inaccuracy at $10^{-18}$ level and their inaccuracy is still improving. Clocks of this accuracy level will be developed for space use. This development is good for laser pulse ranging scheme for Super-ASTROD. The laser pulse timing accuracy of 3 ps is already achieved in T2L2 on board JASON2 satellite.[150] 0.9 mm (3 ps) out of 1300 Gm (8.6 AU) is $7 \times 10^{-16}$. It is comparable to some of the lower frequency strain acceleration noise level. Pulse timing accuracy is still improving. It would be good to study this scheme in more detail.

## Acknowledgements

I would like to thank Wei-Ping Pan for his help in drawing Fig.'s 4-6, and to thank An-Ming Wu and Gang Wang for helpful discussions during many collaboration years. I would also like to thank Science and Technology Commission of Shanghai Municipality (STCSM-14140502500) and Ministry of Science and Technology of China (MOST-2013YQ150829, MOST-2016YFF0101900) for supporting this work in part, and to thank the Kavli Institute for Theoretical Physics, China (KITPC) for funding the Next Detectors for Gravitational Astronomy Program (during the program, the writing of this review was started), and for their hospitality.